\documentclass[english,journal=jctcce,manuscript=article,email=true,etalmode=truncate,maxauthors=0]{achemso}

\setkeys{acs}{
  maxauthors = 6,
  etalmode = firstonly
}

\graphicspath{{/publication_plots/}{publication_plots/}}

  \usepackage{geometry} 
  \usepackage{stackengine}
  \usepackage{titlesec}

  \usepackage[T1]{fontenc} 
  \usepackage{amsmath}     
  \usepackage{amssymb}
  \usepackage{bm}
  \usepackage{mathrsfs}
  \usepackage{etoolbox}
  \usepackage{environ}
  \usepackage[version=4]{mhchem}
  \usepackage{booktabs}
  \usepackage{adjustbox}
  \usepackage{hhline}

  \usepackage{setspace}    
  \usepackage{float}       
  \newfloat{chart}{htbp}{loc}
  \newfloat{graph}{htbp}{loh}
  \newfloat{scheme}{htbp}{los}

  \usepackage{graphicx}    

  \usepackage[english]{babel}
  \usepackage{array}

  \usepackage{subcaption}  
  \usepackage{outlines}    
  \usepackage[normalem]{ulem} 

  \usepackage[numbers,sort&compress,super]{natbib}
  \usepackage{natmove}

  \usepackage{cancel}      
  \usepackage{pdfpages}  
  \usepackage[font=scriptsize,labelfont=bf]{caption}

  \usepackage{hyperref} 
  \usepackage{cleveref}	
  	\crefname{figure}{Figure}{Figures}
  	\crefname{table}{Table}{Tables}
  	\crefname{equation}{Eq.}{Eqs.}
  	\crefname{section}{Section}{Sections}
  	\crefname{subsection}{Section}{Sections}
  	\crefname{subsubsection}{Section}{Sections}
  	\crefname{algorithm}{Algorithm}{Algorithms}

  \usepackage{tikz}

    \usepackage[ruled,vlined]{algorithm2e}

  \usepackage{todonotes}

\usepackage{epstopdf}
\AppendGraphicsExtensions{.tiff}

  \author{Florian Pimpel}
  \affiliation{ Institute for Theoretical Physics,\\ TU Wien, Wiedner Hauptstraße 8-10/136, A-1040\\ Vienna, Austria
}%
     \author{Miguel Morales}
  \affiliation{ Center for Computational Quantum Physics, Flatiron Institute, 162 5th Ave., New York, 10010  NY,  USA}

  \author{Andreas Gr{\"u}neis}
  \affiliation{ Institute for Theoretical Physics,\\ TU Wien, Wiedner Hauptstraße 8-10/136, A-1040\\ Vienna, Austria
}%
  \author{Karl Pierce}
  \affiliation{ The University of Maryland, College Park MD 20742 USA.}
  \email{karl.m.pierce@gmail.com}

  \title{Tensor Hypercontraction of the Coupled Cluster Doubles Amplitudes using a Space-Time Interpolation Grid}

\begin{document}

\begin{abstract}
    We present a tensor hypercontraction (THC) approach for compressing coupled-cluster doubles amplitudes using a joint grid in real space and imaginary time. The proposed method extends interpolative separable density fitting (ISDF) to imaginary-time-propagated orbital-pair densities, enabling the direct construction of a compact, separable representation of the doubles amplitudes without explicitly forming or decomposing the full four-index amplitude tensor. Interpolation points are selected through a pivoted Cholesky decomposition of the space-time pair-density Gram matrix, yielding a basis that is particularly well suited to representing the doubles amplitudes in second-order M{\o}ller-Plesset (MP2). We investigate the accuracy and transferability of this basis within a rank-reduced coupled-cluster singles and doubles (CCSD) framework. Using a THC rank equal to eight times the number of molecular orbitals, CCSD correlation-energy errors remain below $0.1\%$ for all systems considered. Furthermore, we demonstrate that the proposed approximation accurately reproduces relative CCSD energies for both intramolecular rotations and intermolecular dissociation curves, with deviations below 0.1 kcal mol$^{-1}$ at sufficiently large THC ranks. Although the present implementation is designed to assess accuracy rather than computational efficiency, the proposed representation provides a route toward CCSD calculations with quartic computational scaling and offers a promising foundation for extensions to periodic and finite-temperature systems.
\end{abstract}

\section{Introduction}

It is well understood that the accurate evaluation of correlation energy is necessary for the predictive modeling of computational chemistry. 
One of the most successful ways to recover correlation missing from mean-field approaches is through the coupled cluster (CC) family of methods.
Unfortunately, the practical applications of the CC family of methods are limited in part due to the polynomial computational scaling and storage complexity of the approach.
In an effort to combat the computational costs of the methods in molecular systems researchers have been investigating the application of low-rank tensor decompositions. 
Historically, researchers have found the most success in reducing the cost of CC methods by constructing approximations to the two-electron integral (TEI) tensor.
More recently, researchers have started to study the efficient application of low-rank approximations to the $n$-particle excitation tensors.\cite{Hohenstein:2019:JCP,Lesiuk:2020:JCTC,Benedikt:2013:JCP,Schutski:2017:JCP}
Among these attempts is the application of the rank-reduced approximation \cite{Parrish:2019:JCP,Hohenstein:2019:JCP,Lesiuk:2020:JCTC,Hohenstein:2021:JCP,Hohenstein:2022:JCP,Lesiuk:2022:JCP} and the related tensor-structured approach.\cite{Schutski:2017:JCP}
These works have shown that the molecular CC excitation amplitude tensors can be optimized efficiently using an approximate low-rank tensor network format and that this low-rank format can be leveraged to compute accurate chemically interesting energy values with a reduced scaling.

The rank-reduced approximation of the 2-particle excitation tensor can be written as 
\begin{align}\label{eq:rr-ccd}
    t^{ab}_{ij} \approx \sum_{XY} L^{aX}_{i} \zeta^X_Y L^{bY}_{j}
\end{align}
where $L$ are rank-reduced basis transformations and $\zeta$ is the rank-reduced core tensor.
This approximation is practically identical to the one considered by Schutski et al.\cite{Schutski:2017:JCP} in their tensor-structured approximation.
The main difference in the methods, and arguably the biggest flaw in the rank-reduced approximation, is that the rank-reduced approximation assumes that a single good $L$ can be determined efficiently prior to the start of the CC optimization procedure. 
Practically, this allows rank-reduced CC to be more efficient than the tensor-structured approximation because only the rank-reduced core tensor must be determined.
However, methods to determine high-quality $L$ tensors at sufficiently low costs are still an open question.

In their work, Parrish et al. find these $L$ tensors using a least squares loss function 
\begin{align}
    \|t^{ab}_{ij} - \sum_{XY}L^{aX}_i \zeta^X_Y L^{bY}_j\|^2
\end{align}
where $t^{ab}_{ij}$ is the MP2 or MP3 amplitude tensor. 
To solve this problem, the authors employ the eigenvalue decomposition to efficiently compute $L$.
In later work, the authors construct a tensor hyper-contraction (THC)\cite{VRG:hohenstein:2012:JCP,VRG:hohenstein:2012:JCPa,VRG:parrish:2012:JCP,Hohenstein:2013:JCP,VRG:parrish:2014:JCP,Shenvi:2013:JCP,Schutski:2017:JCP,Parrish:2019:JCP,Lee:2019:JCTC,VRG:hummel:2017:JCP,VRG:song::JCP,Hohenstein:2019:JCP,Hohenstein:2021:JCP,Hohenstein:2022:JCP,Jiang:2022:JCTC,Zhao:2023:JCTC,Datar:2024:JCTC} formulation of the wavefunction amplitude tensor by computing the canonical polyadic decomposition of the $L$ tensor.
While methodologically reasonable, this approach can be practically expensive in both memory and time. 
Naively, this strategy requires the construction and decomposition of the order-4 $t$ tensor which carries an $\mathcal{O}(N^4)$ storage cost and $\mathcal{O}(N^5)$ computational cost.

In this work, we aim to develop a method that reliably constructs basis matrices for the THC rank-reduced wavefunction amplitude tensor at lower computational cost.
To achieve this objective, we develop a novel strategy 
that leverages the machinery of the interpolative separable density fitting (ISDF) decomposition.\cite{Lee:2019:JCTC, VRG:hu:2017:JCTC,VRG:lu:2015:JCP,ISDF_GW_Duchemin21,ISDF_RPA_Yeh2024,ISDF_GW_Yeh2024,ISDF_AFQMC_Malone19}
Specifically, our method employs a subset selection algorithm to construct a space--time interpolation basis for the time-dependent orbital-pair densities.
We demonstrate the accuracy of this interpolation scheme by constructing a THC-like space--time approximation to the MP1 wavefunction. Finally, we discuss how and why this interpolative basis can be incorporated into the rank-reduced CC optimization framework.

This manuscript is organized as follows.
First, in \cref{sec:formal} we introduce the standard tensor decomposition strategies used in the rank-reduced CC method. In \cref{sec:t-isdf} we describe the novel imaginary-time dependent interpolation procedure.
Next we recap the CC method, describe the rank-reduced coupled cluster method and motivate the use of our novel interpolation basis in the rank-reduced procedure.
In \cref{sec:details} we describe the details of our computational experiments and in \cref{sec:results} we illustrate that the interpolated rank-reduced CC method is reliably accurate for molecular calculations.
Finally in \cref{sec:conclusions} we summarize the work constructed here, discuss further applications of the rank-reduced formalism and discuss future work.

\section{Formalism}\label{sec:formal}
\subsection{Low-Rank Tensor Approximations}
In this section we discuss the relevant low-rank tensor decompositions that we take advantage of in this work.

In electronic structure theory, low-rank decompositions of the four-center two-electron integral (TEI) tensor have proven particularly effective in reducing computational scaling.
The TEI tensor can be written in an arbitrary set of basis functions $\{\phi\}$ as
\begin{align}\label{eq:TEI}
    g^{\mu\nu}_{\kappa\sigma} &= \iint \phi^*_\mu(r_1) \phi^*_\nu(r_2) \hat{g}(r_1, r_2)\phi_\kappa(r_1) \phi_\sigma(r_2) dr_1 dr_2 \nonumber \\
    &= \iint \rho_{\mu\kappa}(r_1) \hat{g}(r_1, r_2)  \rho_{\nu\sigma}(r_2) dr_1 dr_2,
\end{align}
where we call $\rho_{\mu\kappa}(r) = \phi^*_\mu(r) \phi_\kappa(r)$ the orbital-pair density. The kernel $\hat{g}(r_1, r_2)$ can be any positive definite function, however,
in this work we only consider this kernel to be the Coulomb interaction kernel, $\hat{g}(r_1, r_2) = |r_1 - r_2|^{-1}$.
The most popular decomposition strategy for the TEI tensor is the density fitting (DF) approximation.\cite{VRG:whitten:1973:JCP,VRG:dunlap:1979:JCP,Vahtras:1993:CPL,VRG:jung:2005:PNAS,VRG:mintmire:1982:PRA}
The DF approximation first recasts the order-4 TEI tensor as a square, symmetric matrix, by fusing the two indices that represent the same particle index.
Subsequently, one can form the DF approximation as the square-root of the TEI matrix
\begin{align}
    g^{\mu\nu}_{\kappa\sigma} \overset{\mathrm{DF}}{\approx} \sum_{X} B^{\mu X}_{\kappa} B^{\nu X}_{\sigma}
\end{align}
where $X$ is the rank of the matricized tensor and is referred to as the auxiliary basis set and $B$ are the DF-approximated TEI tensors.

The THC approximation further reduces the dimensionality of the DF-approximated TEI tensor from $\mathcal{O}(N^3)$ to $\mathcal{O}(N^2)$ by casting the order-4 TEI as a network of 5 matrices
\begin{align}\label{eq:thc_tei}
    g^{\mu\nu}_{\kappa\sigma} \overset{\mathrm{THC}}\approx\sum_{PQ} U^\mu_P U^P_\kappa Z^P_Q U^\nu_Q U^Q_\sigma.
\end{align}
In \cref{eq:thc_tei}, the tensors $U$ are denoted as the THC factor matrices and $Z$ is the THC core matrix.
Please note, in this study we only consider decompositions where $P=Q=R_\mathrm{THC}$ and we denote $R_\mathrm{THC}$ as the THC rank.
One can construct a three center basis transformation tensor from these THC factor matrices by leveraging the Khatri-Rao tensor product, i.e.
\begin{align}
    M^{\mu P}_{\kappa} = U^{\mu}_{P} U^{P}_{\kappa}.
\end{align}
Because of this tensor product relationship, the THC rank may be referred to as a 'hyper-index' and, unlike Einstein summation notation, more than 2 tensors may carry a hyper-index.
One can see that, like the DF approximation, the THC requires that the two indices that represent a single particle be bound to the same basis transformation tensor and, therefore, the same THC rank index.
For example, here the indices $\mu$ and $\kappa$ are representative of particle-1 and are therefore pinned to the $P$ index.
Furthermore, if we assume that the basis functions denoted by $\mu$, $\nu$, $\kappa$ and $\sigma$ all span the same functional subspace, then the THC factor matrices are all exactly equivalent, i.e. $U^\mu_p = (U^P_\kappa)^H = U^{\nu}_Q = (U^Q_\sigma)^H$ where $H$ denotes the Hermitian transpose.

\subsection{Interpolative Separable Density Fitting}

The pseudospectral approach\cite{VRG:friesner:1985:CPL,Friesner:1986:JCP,Friesner:1991:ARPC,VRG:ringnalda:1990:JCP,Greeley:1994:JCP} first introduced a real-space grid approximation strategy for the TEI tensor. 
In this decomposition, the integrals over the real-space position of a single particle are replaced with a quadrature approximation
\begin{align}
    g^{\mu\nu}_{\kappa\sigma} \approx \sum_{r_g}w_g\phi_{\mu}^*(r_g)\phi_\kappa(r_g) \int g(r_g,r_2) \phi_\nu^*(r_2) \phi_\sigma(r_2) dr_2
\end{align}
where $r_g$ is some set of quadrature grid points in real space.
This approximation allows one to express the order-4 tensor as a product of three tensors
\begin{align}
    g^{\mu\nu}_{\kappa\sigma} \approx X^{\mu}_g X^g_{\kappa} Y^{\nu g}_{\sigma}
\end{align}
where $X^{\mu}_g = \sqrt w_g \phi_\mu(r_g)$ and $Y^{\nu g}_{\sigma} = \int \hat{g}(r_g,r_2) \phi^*_\nu(r_2) \phi_\sigma(r_2)dr_2$.
The THC approximation extended this idea by replacing both integrals in the TEI expression with quadrature approximations.
The issue with this grid-based approach is that the kernel $\hat{g}(r_{g_1}, r_{g_2}) = \frac{1}{r_{g_1} - r_{g_2}}$ (where $r_{g_1}$ is the quadrature grid for the density associated with particle 1 and $r_{g_2}$ is the quadrature grid associated with particle 2) diverges when $r_{g_1} = r_{g_2}$.
One way to rectify this issue is to approximate the TEI matrix by solving the following least squares equation 
\begin{align}
    \min_Z \| g^{\mu\nu}_{\kappa\sigma} - \sum_{PQ} X^\mu_P X^P_\kappa Z^P_Q X^\nu_Q X^Q_\sigma\|
\end{align}
where $P$ and $Q$ label particle 1 and 2 quadrature grid points respectively.

Within this formalism, interpolative separable density fitting (ISDF) provides a means of selecting real-space interpolation points for the PS or THC representations. The real-space grid points are selected by minimizing the error in an interpolative approximation of the orbital-pair densities
\begin{align}
    \rho_{\mu\kappa}(r) = \phi_\mu^*(r) \phi_\kappa(r) .
\end{align}
Although these densities are continuous functions over the space domain, the densities can be accurately discretized using an extremely tight grid basis \(\{r_g\}_{g=1}^{N_g}\).
The interpolative decomposition seeks to select \(N_p\ll N_g\) of the points in real space such that
\begin{align}
    \| \rho_{\mu\kappa}(r_g) - C  C^+ \rho_{\mu\kappa}(r_g)  \| < \epsilon
\end{align}
where $C^{r_p}_{\mu\kappa} = \sum_{r_g}\rho_{\mu\kappa}(r_g) P^{r_g}_{r_p}$, $P$ is an incomplete, pivoted identity matrix and \(C^+\) denotes the Moore–Penrose pseudoinverse.
There are many ways to determine the $P$ matrix, one way is to leverage the square, positive-semidefinite Gram matrix
\begin{align}
    K(r_g, r_g') = \sum_{\mu\kappa}\rho^*_{\mu\kappa}(r_g) \rho_{\mu\kappa}(r_g').
\end{align}
Greedy column selection can be implemented through a pivoted Cholesky factorization of \(K\), selecting, at each step, the largest diagonal entry of the residual matrix. Although storing the full \(N_g\times N_g\) Gram matrix may be prohibitive, this procedure requires only the diagonal elements of the full Gram matrix and the selected columns,
both of which can be evaluated on the fly using orbital values from standard integral libraries.

\subsection{Imaginary-Time resolved Interpolation Basis for the Orbital-Pair Densities} \label{sec:t-isdf}

In this work, we extend this theory to imaginary-time-propagated pair densities.
In particular, we are motivated by the Laplace-transform approximated amplitude tensor (MP1), used to compute the second-order M{\o}ller-Plesset energy (MP2).
This tensor can be written as 
\begin{align}\label{eq:mp1_wfn}
    t^{ab}_{ij} =- g^{ab}_{ij} * [D^{-1}]^{ab}_{ij} 
\end{align}
where $*$ is the element-wise, Hadamard product.
The energy denominator tensor can be expressed using the Laplace transformation which can also be approximated using a quadrature approximation
\begin{align}\label{eq:LT_quad}
     [D^{-1}]^{ab}_{ij}  = \frac{1}{\epsilon_a +\epsilon_b - \epsilon_i - \epsilon_j} \overset{\mathrm{LT}}{=} \int_0^\infty e^{-D^{ab}_{ij}t} dt 
     \approx \sum_{\ell}^\alpha w_\ell e^{-D^{ab}_{ij}t_\ell}
\end{align}
where $\alpha$ is the number of imaginary-time quadrature points.
Substituting this Laplace transform in \cref{eq:mp1_wfn} provides
\begin{align}
    t^{ab}_{ij} = -g^{ab}_{ij} \int_0^\infty e^{-D^{ab}_{ij}t} dt.
\end{align}
By expanding $g$ in terms of the AO to MO coefficients and its integral operator we can rewrite the MP1 wavefunction as
\begin{align}
    t^{ab}_{ij} = -\sum_{\mu\nu\kappa\sigma} [\iint \phi^*_\mu(r_1) \phi^*_\nu(r_2) \hat{g}(r_1, r_2)\phi_\kappa(r_1) \phi_\sigma(r_2) dr_1 dr_2] {C^*}^{a}_\mu {C^*}^{b}_{\nu} C^\kappa_i C^\sigma_j [\int e^{-D^{ab}_{ij}t}dt].
\end{align}
By leveraging the separability of the exponential, we can rewrite the expression as
\begin{align}
    t^{ab}_{ij} = -\sum_{\mu\nu\kappa\sigma} [\iint \phi^*_\mu(r_1) \phi^*_\nu(r_2) \hat{g}(r_1, r_2)\phi_\kappa(r_1) \phi_\sigma(r_2) dr_1 dr_2] [\int ({C^*}^{a}_\mu e^{-\epsilon_at} ) ({C^*}^{b}_{\nu} e^{-\epsilon_b t}) (C^\kappa_i e^{\epsilon_i t}) (C^\sigma_j e^{\epsilon_j t})  dt].
\end{align}
Next, we can combine all the integrals to find 
\begin{align}
    t^{ab}_{ij} = -\sum_{\mu\nu\kappa\sigma} \iiint (\phi^*_\mu(r_1){C^*}^{a}_\mu e^{-\epsilon_at} )
    (\phi^*_\nu(r_2) {C^*}^{b}_{\nu} e^{-\epsilon_b t})
    \hat{g}(r_1, r_2)
    (\phi_\kappa(r_1) C^\kappa_i e^{\epsilon_i t})
    (\phi_\sigma(r_2) C^\sigma_j e^{\epsilon_j t})
    dr_1 dr_2 dt.
\end{align}
With this we can identify an imaginary-time-dependent orbital-pair density 
\begin{align}\label{eq:td-density}
    \rho_{ai}(r_1, t) = [\sum_{\mu} \phi^*_\mu(r_1){C^*}^a_\mu e^{-\epsilon_a t}][\sum_\kappa \phi_\kappa(r_1)C^\kappa_i e^{\epsilon_i t}] = \phi_a^*(r_1,t) \phi_i(r_1,t).
\end{align}
With this object, we may now introduce an interpolation point selection procedure similar to the ISDF method.
By representing both $r_1$ and $t$ on a quadrature grid, i.e. $r_g$ and $t_\ell$, we can jointly select interpolation points in space and imaginary time.
To determine the space-time interpolation matrix, we can again form a new space-time Gram matrix
\begin{align}
    \hat{K}^{r_g t_\ell}_{r_g' t_\ell'} = \sum_{ai} \rho^*_{ai}(r_g,t_\ell) \rho_{ai}(r_g',t_\ell').
    \label{eq:gram_st}
\end{align}
The resulting space-time Gram matrix is $\alpha^2$ times larger than the original Gram matrix of the non-propagated pair densities. 
However, by leveraging the definition of the time-dependent density, \cref{eq:td-density}, one can construct a matrix-free algorithm to efficiently determine the interpolation factorization, $\hat{C}^{S}_{ai} = \sum_{r_g t_\ell}\rho_{ai}(r_g,t_\ell) P^{r_gt_\ell}_{S}$.
We will denote this new mixed time-space quadrature basis with the labels $S$ and $T$.
The interpolation coefficients retain the following separable structure:
\begin{align}
    \hat{C}^S_{ai} = \sum_{r_gt_\ell} \rho_{ai}(r_g,t_\ell) P^{r_gt_\ell}_{S} = [\sum_{r_gt_\ell} \phi_a^*(r_g,t_\ell)P^{r_gt_\ell}_{S}] \odot [\sum_{r_gt_\ell} \phi_i(r_g,t_\ell)P^{r_gt_\ell}_S] = V^S_a\odot V^S_i
\end{align}
where $\odot$ is the Khatri-Rao product.
Once the grid has been selected, this new basis can be introduced into the rank-reduced CC procedure (discussed below) or, in the case of MP2,
used to determine the amplitude's core tensor in the interpolation basis using the least-squares problem
\begin{align}\label{eq:mp2_loss}
    \min_\zeta \| t^{ab}_{ij} - V^{a}_S V^S_i \zeta^S_T V^b_T V^T_j \|^2_2 = \min_\zeta \|-\frac{g^{ab}_{ij}}{D^{ab}_{ij}} - V^{a}_S V^S_i \zeta^S_T V^b_T V^T_j \|^2_2.
\end{align}

\begin{figure}[htbp]
  \centering
  \includegraphics[width=0.88\textwidth]{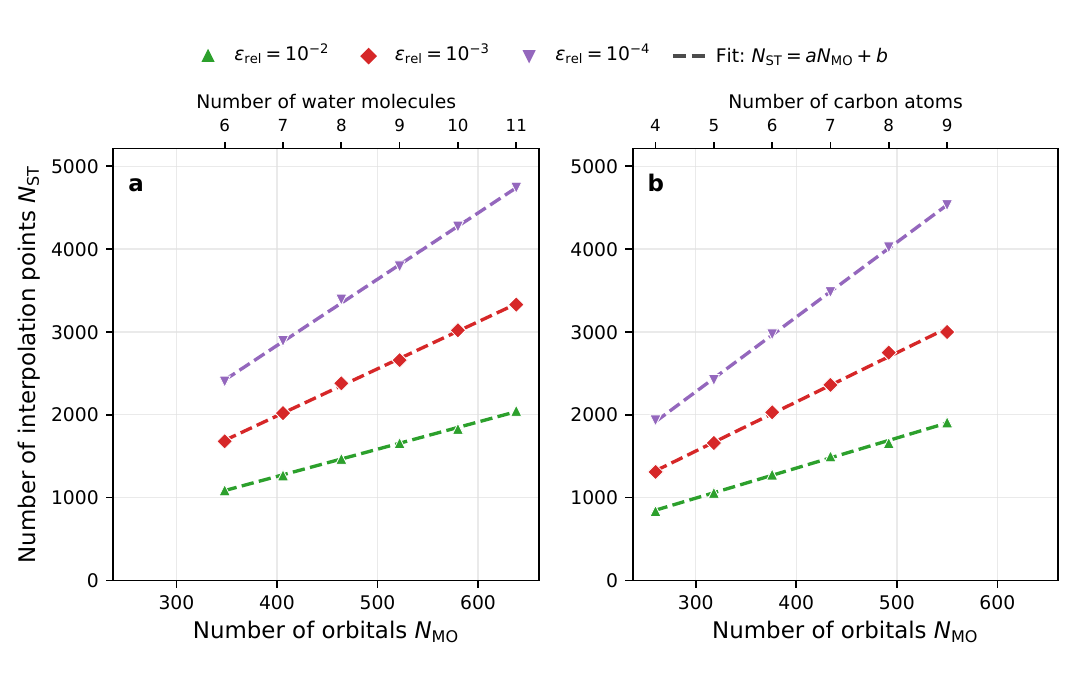}
  \caption{Number of space-time interpolation points required to approximate the doubles amplitudes to a specified accuracy of the DF-MP2 energy for water clusters (a) and linear alkanes (b). Points are added until $\| E_\mathrm{corr} - E^\mathrm{THC}_\mathrm{corr}\| / \|E_\mathrm{corr}\| \leq \varepsilon_\mathrm{rel}$. Dashed lines represent a linear least squares fitting for the data points of each accuracy threshold. The orbital basis set is the cc-pVTZ basis and the density fitting auxiliary basis set is the associated cc-pVTZ-RI basis. }
  \label{fig:mp2-rank}
\end{figure}
\noindent Note that this core-finding step is specific to MP2, and not part of the construction of the interpolation basis, i.e. if the aim is to use this basis to approximate a different tensor, this step can be skipped.

To assess the compactness of the space-time interpolation basis, we examine the number of interpolation points required to approximate the MP1 doubles amplitudes such that the resulting MP2 correlation energy meets a prescribed accuracy.
We compute the amplitude-approximated, restricted MP2 energy as
\begin{align}
    E^{\mathrm{THC}}_{\mathrm{corr}} = \sum_{abij} 2 \tilde{t}^{ab}_{ij} \bar{g}^{ab}_{ij} - \sum_{abij} \tilde{t}^{ab}_{ij} \bar{g}^{ba}_{ij}
\end{align}
where $\tilde{t}^{ab}_{ij} = \sum_{ST} V^a_S V^S_i \zeta^S_T V^b_T V^T_j$ and $\bar{g}^{ab}_{ij} = \sum_X B^{aX}_i B^{bX}_j$ is the density fitting approximation.
For this experiment, we use the DF-approximated restricted MP2 energy as the reference energy.
\cref{fig:mp2-rank} shows that for both water clusters and linear alkanes, the required interpolation rank grows approximately linearly with system size over the range studied. This behavior indicates that the amplitude representation can be systematically extended to larger systems while retaining a basis whose size remains proportional to system size. The accuracy criterion is imposed on the correlation energy rather than directly on the amplitudes:
\begin{align}
    \frac{\| E_\mathrm{corr} - E^\mathrm{THC}_\mathrm{corr}\|} { \|E_\mathrm{corr}\|} \leq \varepsilon_\mathrm{rel}
\end{align}

\subsection{Coupled Cluster Theory and Optimization}
The goal of the CC approach is to construct an $N$-particle wave function that recovers the ‘dynamical correlation’ which is missing from the Slater determinant picture of mean-field electronic structure methods.\cite{VRG:cizek:1966:JCP, Cizek:1969:ACP,Cizek:1971:IJQC,VRG:crawford:2000:RCC,VRG:tajti:2004:JCP,VRG:harding:2008:JCP,Thorpe:2019:JCP,VRG:purvis:1982:JCP}
The total wavefunction can be defined as 
\begin{align}\label{eq:cc_wfn}
   |\Psi_\mathrm{CC} \rangle = e^{T} | \Phi_0 \rangle
\end{align}
where $\Phi_0$ is a ground-state wavefunction approximated by a single set of antisymmetrized one particle functions and $\Psi_\mathrm{CC}$
is a sum of $n$-particle wavefunctions built from the set of single-particle functions.
Here $T$ is the excitation operator
\begin{align}
    T = T_1 + T_2 + \cdots
\end{align}
such that $T_n$ is a $n$-particle excitation operator
\begin{align}
    T_n = \sum_{i,j,...,a,b,...} t^{ab...}_{ij...} a^{ij...}_{ab...}.
\end{align}
In this equation, the labels $i,j,k,...$ are representative of the set of $O$ occupied orbital basis functions and the labels $a,b,c,...$ are representative of the set of $V$ unoccupied orbital basis functions.
The goal of the CC method is to optimize the set of excitation coefficient tensors for a given value of $n$.

In this work, we study only the coupled cluster with single and double excitations (CCSD) method, i.e. the excitation operator $T$ is truncated at $n=2$.
The CCSD method seeks to find the optimal amplitudes of the single and double excitations using a coupled set of nonlinear equations, 
\begin{align}\label{eq:ccsd}
    0 &= \langle \Phi^{a}_{i} | \bar{H} | \Phi_0 \rangle \\ \nonumber
    0 &= \langle \Phi^{ab}_{ij} | \bar{H} | \Phi_0 \rangle
\end{align}
where $\bar{H}$ is the similarity transformed Hamiltonian, which will not be discussed here.
In \cref{eq:ccsd}, 
$\Phi^a_i$ represents the excited wavefunction obtained from swapping of the $i$th occupied orbital function in the Slater determinant ground state, $\Phi_0$, with the $a$th unoccupied orbital function and doubly excited wavefunction, $\Phi^{ab}_{ij}$, is determined in a similar manner.
The optimized CCSD amplitudes are stored in the order-2 $t^a_i$ and order-4 $t^{ab}_{ij}$ tensors, respectively.
And, furthermore, they are optimized using a standard nonlinear gradient descent optimization technique.
Though the Coupled Cluster methods find great success in recovering the dynamical correlation, the family is plagued by a restrictively large computational complexity. 
In the following sections we introduce useful tensor decomposition strategies and employ the rank-reduced approximation to reduce the computational scaling of the CCSD method.

\subsection{The Rank-Reduced Coupled Cluster Method} \label{sec:rr-ccsd}

In its original formulation\cite{Parrish:2019:JCP}, the rank-reduced approximation decomposes both the 1- and 2-particle excitation tensors as 
\begin{align}\label{eq:rrt1}
    t^a_i \approx \sum_{V} L^{aV}_{i} \zeta^V
\end{align}
and 
\begin{align}\label{eq:rr-t2}
    t^{ab}_{ij} \approx \sum_{XY} L^{aX}_{i} \zeta^X_Y L^{bY}_{j}.
\end{align}
In this work we do not consider the rank-reduced approximation of the 1-particle excitation tensor in \cref{eq:rrt1}.
As pointed out previously, the rank-reduced formulation in \cref{eq:rr-t2} was, later, further decomposed to give the 2-particle excitation tensor a THC-like approximation \cite{Hohenstein:2022:JCP}
\begin{align}\label{eq:thc-cc-r2}
t^{ab}_{ij} \approx \sum_{ST} V^a_S V^S_i \zeta^S_T V^{b}_{T} V^{T}_{j}.
\end{align}
In equation \cref{eq:rr-t2} and \cref{eq:thc-cc-r2}, we denote the tensors $L$ and $V$ as the rank-reduced transformation basis tensors.

The THC approximation in \cref{eq:thc-cc-r2} can be determined from \cref{eq:rr-t2} by computing the canonical polyadic decomposition\cite{VRG:carroll:1970:P,VRG:Harshman:1970:WPP} of each of the three-index rank-reduced transformation basis
\begin{align}
    L^{aX}_{i} \approx \sum_{S} V^a_S V^S_i Q^X_S.
\end{align}
Using this decomposition, we can rewrite \cref{eq:rr-t2}
\begin{align}
    t^{ab}_{ij} \approx \sum_{XY} L^{aX}_{i} \zeta^X_Y L^{bY}_{j} \approx \sum_{ST} V^a_S V^S_i (\sum_{XY} Q^X_S \zeta^X_Y Q^Y_T) V^b_T V^T_j
\end{align}
and by summing over $X$ and $Y$ we find $\zeta^S_T = \sum_{XY} Q^X_S \zeta^X_Y Q^Y_T$.
In practice, this transformation is not necessary as either \cref{eq:rr-t2} or \cref{eq:thc-cc-r2} is proposed and the core tensor, $\zeta$ is dynamically determined using the CCSD optimization loop.
As pointed out previously, the rank-reduced CCSD procedure assumes that the transformation basis (either $L$ or $V$) is known and fixed prior to the CCSD optimization procedure. Parrish et al. calculate this basis by decomposing an approximate doubles amplitude tensor.
The easiest choice is the MP1 amplitude tensor, since it is the first order approximation to the CCSD amplitude tensor.
As a result, the rank-reduced CCSD optimization only optimizes the core matrix $\zeta$ when solving \cref{eq:ccsd}.

In canonical CCSD, the doubles amplitude tensor is 
iteratively optimized using the residual 
$\langle \phi^{ab}_{ij} | \bar{H} | \phi_0\rangle$.
The residual equation may be rearranged as
\begin{align}\label{eq:ccsd_doubles}
0 = \langle \phi^{ab}_{ij} | \bar{H} | \phi_0\rangle = D^{ab}_{ij} t^{ab}_{ij} + R^{ab}_{ij},
\end{align}
where  $D^{ab}_{ij} = \epsilon_a + \epsilon_b - \epsilon_i - \epsilon_j$ is
the orbital-energy tensor expressed in canonical molecular orbital basis. 
Here, \(R_{ij}^{ab}\) denotes the remaining
right-hand side of the residual equation after the diagonal contribution
\(D_{ij}^{ab}t_{ij}^{ab}\) has been separated from the residual.
This form provides us with a Jacobi update to the doubles amplitude tensor
\begin{align}\label{eq:full_t_update}
    t^{ab}_{ij} =-\frac{ R^{ab}_{ij}}{ D^{ab}_{ij} }.
\end{align}.

To obtain a similar optimization strategy for the rank reduced core matrix $\zeta$, Parrish et al. project \cref{eq:ccsd_doubles} using the chosen subspace of the doubles amplitude tensor
\begin{align}\label{eq:ccsd_proj}
    0 =  \sum_{abij} \langle \Phi^{ab}_{ij} | \bar{H} | \Phi_0 \rangle L^{aP}_i L^{bQ}_j =
    \sum_{abij} L^{aP}_i \left( D^a_i t^{ab}_{ij} + 
    t^{ab}_{ij} D^b_j \right) L^{bQ}_j + \sum_{abij} L^{aP}_i R^{ab}_{ij} L^{bQ}_j.
\end{align}
We point out that this approach effectively restricts the de-excitation amplitude tensor in the Lagrangian formulation to live in the same subspace
as the doubles amplitude tensor.\cite{Parrish:2019:JCP}
Next, Parrish et al. substitute the rank-reduced approximation of the doubles amplitude into \cref{eq:ccsd_proj}:
\begin{align}
    0 &= \sum_{abijXY} L^{aP}_{i} [D^a_i L^{aX}_{i} \zeta^X_Y L^{bY}_{j} + L^{aX}_{i} \zeta^X_Y L^{bY}_{j}D^{b}_{j}] L^{bQ}_{j} + \sum_{abij} L^{aP}_{i}R^{ab}_{ij}L^{bQ}_{j} \\
    & = \sum_{XY}A^P_X \zeta^X_Y B^Y_Q + \sum_{XY} B^P_X \zeta^X_Y A^Y_Q + R^P_Q. \nonumber
\end{align}
The rank-reduced core matrix $\zeta$ can now be solved in this projected basis, using this Sylvester equation. With this formulation,
together with the density fitting approximation, Parrish et al. are able to reduce the computational cost of CCSD. \cite{Parrish:2019:JCP}
By adding two additional approximations in the evaluation of the residual, Lesiuk reduces the computational scaling of CCSD from
$\mathcal{O}(N^6)$ to $\mathcal{O}(N^5)$.\cite{Lesiuk:2022:JCP} If one uses the THC approximation (\cref{eq:thc-cc-r2}) for both the 2-particle excitation tensor and the THC approximation of the TEI tensor (\cref{eq:thc_tei}) then one can further reduce the complexity of CCSD to $\mathcal{O}(N^4)$. \cite{Hohenstein:2022:JCP}
Note that this cost reduction comes from two aspects in the rank-reduced formulation -- the low-rank representation of the doubles amplitude tensor,
and the computation of only the projected right-hand side of the residual equation $R^P_Q$ instead of the full $R^{ab}_{ij}$.
This can be done efficiently by distributing the contraction over $a,b,i$ and $j$ over all the terms in the CCSD residual.

As discussed above, the doubles amplitudes used by MP2 constitute the first-order contribution to the CCSD doubles amplitudes, and compression bases derived from them have proven effective in rank-reduced CCSD for a wide range of systems.\cite{Parrish:2019:JCP,Lesiuk:2022:JCP} Since the proposed imaginary-time-resolved interpolation basis is well suited to compressing MP2 amplitudes, we investigate its use for representing the CCSD doubles amplitudes within the same formalism.

\section{Computational Details}\label{sec:details}

For each molecular geometry, we performed a neutral, closed-shell,
density-fitted restricted Hartree--Fock calculation with PySCF \cite{sun_pyscf_2018,sun_python_2026} in the
cc-pVTZ orbital basis \cite{Dunning1989} and the cc-pVTZ-RI \cite{Weigend2002a} auxiliary basis. PySCF's standard
chemical-core selection was used to freeze core orbitals in both conventional and
rank-constrained correlation calculations.  The SCF energy convergence
threshold 
was
$10^{-10}$~Hartree. In these scans, conventional and rank-constrained CCSD
calculations used an energy threshold of
$10^{-8}$~Hartree, an amplitude-norm threshold of $10^{-6}$, and six DIIS vectors.

Water clusters from the TIP4P dataset\cite{Wales:1998:CPL}  and linear alkanes constructed in Avogadro2\cite{hanwell_m_hutchison_g_avogadro_2026} provide a systematic size series with
different chemistry: the former probe hydrogen-bonded environments, while
the latter lengthen a saturated hydrocarbon chain. Like the MP2 analysis provided earlier, these molecules allow us to examine
the interpolation-point count required at
fixed relative accuracy of the total correlation energy as $N_{\mathrm{MO}}$ grows.
Next, we assessed relative energy accuracies by analyzing the rotational energies of butane central C--C at various torsion angles. We used 25 angles from
$0^\circ$ to $360^\circ$ in $15^\circ$ steps. Geometries at each angle were optimized
at the MP2/cc-pVDZ level using PySCF.
Furthermore, a complementary intermolecular energy test considered the dissociation curve of eight small molecule dimers from a list of biologically important weakly bound systems, the S66$\times$8 dataset\footnote{The S66$\times$8 consists of (1) Water--water, (2) water--methanol, (3)
water--methylamine, (4) methylamine--methanol, (5) Ethyne--water, (6) benzene--ethyne, (7) benzene--ethene and (8)
methylamine--pyridine}, at eight intermolecular distances.
The intermolecular distance ranged from
$R/R_{\mathrm e}=0.90$ to $2.00$, where $R_{\mathrm e}$ is the equilibrium distance.
Geometries of the S66$\times$8 were taken from BEGDB \cite{Rezac:2008:CCCC}.

For each dimer or butane
scan, we compare relative total-energy profiles obtained by subtracting
each method's energy at $R/R_{\mathrm e}=2$ or $0^\circ$, respectively.
This tests the shape of the energy surface while cancelling a constant
energy offset; these curves are not monomer-subtracted interaction energies.

\subsection{Construction of the space-time-interpolation basis}

For each system, we constructed an atom-centered Becke quadrature grid \cite{becke_multicenter_1988}
with PySCF's molecular DFT grid builder at grid level~5. We retained every
grid coordinate $\mathbf r_g$ and evaluated all
atomic-orbital (AO) basis functions $\chi_\mu(\mathbf r_m)$ at those
coordinates with PySCF's AO evaluator. The Becke quadrature weight
was not used in this procedure. The AO values were transformed with
the RHF coefficient matrix $C^a_{\mu}$ and $C^{\mu}_i$ to obtain
the values of the molecular orbitals on the grid $\phi_{a/i}(\mathbf r_g)$.

For each geometry, we generated a 15-point minimax Laplace quadrature from
the lowest and highest occupied and virtual orbital energies \cite{Takatsuka:2008:JCP,helmich-paris_improvements_2016}. We use the exponents
$t_\ell$ and weights $w_\ell$, with $\ell=1,\ldots,15$, to define the raw occupied
and virtual THC factors
\begin{equation}
 V^{r_g\ell }_i=w_\ell^{1/4}e^{t_\ell\varepsilon_i}\phi_i( r_g),
 \qquad
 V^a_{r_g\ell}=w_\ell^{1/4}e^{-t_\ell\varepsilon_a}\phi_a( r_g).
\end{equation}
We prune the set of compound indices  $(r_g,\ell)$ by using a maximum-diagonal pivoted Cholesky procedure on the
positive-semidefinite Gram matrix $S=\rho\rho^\dagger$, given in \cref{eq:gram_st}. 
A fixed rank was requested, smaller than the occupied--virtual
product-space dimension $N_{\mathrm{occ}}N_{\mathrm{vir}}$ as the stopping criterion.
For the larger systems, we used a randomized approach to preselect a pool 
of candidate indices and validated the selected span against the full grid in blocks.
Analysis of this procedure will be part of future work.

\subsection{Rank-reduced CCSD}

For CCSD calculations we expanded the doubles amplitudes in
the imaginary-time resolved interpolation basis. Numerically
dependent pair vectors of $U_{ia,P}=X^{(o)}_{P i}X^{(v)}_{P a}$ were removed with a relative eigenvalue cutoff of
$10^{-10}$ applied to the column-normalized pair metric. We define the
relative THC rank as $\alpha=N_{\mathrm{ST}}/N_{\mathrm{MO}}$, where
$N_{\mathrm{ST}}$ counts selected compound interpolation points before this
numerical reduction and $N_{\mathrm{MO}}$ is the total number of molecular
orbitals.

We used the rank-reduced CCSD method, introduced in \cref{sec:rr-ccsd},
to determine the rank-reduced amplitudes in the selected
occupied--virtual pair basis. 
For the scope of this work, we did not implement a low-scaling
algorithm, but simulated the rank-reduced equations in order to assess
its accuracy. At each iteration we evaluated the
ordinary restricted CCSD residual using a reconstructed doubles tensor,
then projected the doubles numerator into the amplitude basis, and solved
the projected Sylvester equation. The conventional density-fitted CCSD calculation
is used as a reference and
uses the same orbitals, frozen space, and integral backend as the
rank-constrained calculation.
We leave the presentation of an implementation that fully leverages the low-scaling complexity accessible by the outlined low-rank representation to future work.

\section{Results}\label{sec:results}

In this section, we assess the accuracy and robustness of rank-reduced coupled cluster singles and doubles using the proposed imaginary-time resolved interpolation basis of the pair densities.
In \cref{sec:tot_energies}, we determine the size of the space-time grid required to achieve a specified target accuracy by calculating the total energies of systems of different types and sizes. Finally, in \cref{sec:rel_energies}, we quantify the errors introduced by residual compression in CCSD energy differences for a range of applications.

\subsection{Accuracy of Total Energies}\label{sec:tot_energies}

We want to assess the accuracy of total CCSD energies obtained in the imaginary-time resolved pair-density basis and roughly
estimate the THC basis size $N_\mathrm{ST}$ required to achieve a given accuracy.
As an initial test, we calculate the CCSD correlation energies of water clusters containing four, six, eight
and ten water molecules and of linear alkanes containing four, six, and eight carbon atoms.
Because the Hartree–Fock contribution is unaffected by the residual approximation, the error in the correlation energy is equal to the error in the total energy.

\begin{figure}[htbp]
  \centering
  \includegraphics[width=0.92\textwidth]{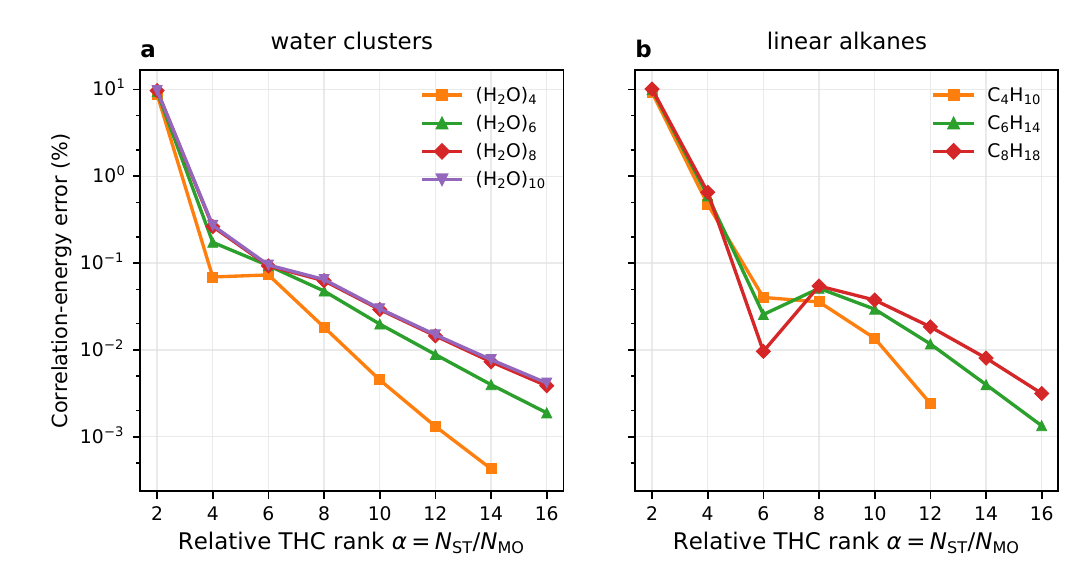}
  \caption{Absolute error in the total CCSD correlation energy in percent as a function
  of $\alpha=N_\mathrm{ST}/N_{\mathrm{MO}}$ for water clusters (a) and linear alkanes (b).}
  \label{fig:size-convergence}
\end{figure}

Figure \ref{fig:size-convergence} shows the error of the space-time grid approximated method relative to the reference CCSD calculation

\begin{equation}
\varepsilon_{\mathrm{corr}}
=
\frac{
\left|E_{\mathrm{corr}}^{\mathrm{THC}}
-
E_{\mathrm{corr}}^{\mathrm{ref}}\right|
}{
\left|E_{\mathrm{corr}}^{\mathrm{ref}}\right|
},
\label{eq:corr-energy-error}
\end{equation}

as a function of the THC rank. The THC auxiliary dimension is expressed as a multiple of the number of molecular orbitals,
\[
N_\mathrm{ST} = \alpha N_{\mathrm{MO}}.
\]
For sufficiently large systems, we expect the value of $\alpha$ required to achieve a fixed accuracy to become independent of system size.

Figure \ref{fig:size-convergence} shows that a relative THC basis size of $\alpha=8$ yields energy errors below $0.1\%$ of the reference energy for all systems considered. Two convergence regimes are apparent: an initial rapid decrease in the error up to $\alpha\approx 4$–$6$, followed by slower convergence. The transition between these regimes is accompanied by a change in the sign of the error. The relative-error curves lie close together in the initial regime, whereas in the slower regime, approximate agreement between the curves appears to require somewhat larger systems.

The two convergence regimes are consistent with the use of an imaginary-time-resolved pair-density basis tailored to the MP2 doubles amplitudes. The MP2 contribution is the leading-order term in the CCSD doubles amplitudes and is expected to dominate the error at small basis sizes. Its efficient representation can therefore explain the rapid initial convergence. At larger basis sizes, the remaining error increasingly reflects contributions beyond MP2, which are less efficiently represented by this basis.

The compressibility of the contributions to the CCSD doubles amplitudes can be further investigated by considering different terms in the amplitude equations. For simplicity, we restrict the analysis to doubles excitations (CCD) and examine selected contributions to the restricted CCD amplitude equations. Retaining only the diagonal Fock matrix contributions and the electron repulsion integrals yields the doubles amplitude equations for MP2. The direct-ring CCD (drCCD) approximation additionally includes particle-hole-bubble contributions while neglecting particle exchange.\cite{scuseria_ground_2008,shepherd_coupled_2014} The direct contribution to the drCCD correlation energy is equivalent to the direct random-phase approximation (RPA), while the exchange contribution is commonly referred to as second-order screened exchange (SOSEX).\cite{gruneis_making_2009}

\begin{figure}[htbp]
  \centering
  \includegraphics[width=0.92\textwidth]{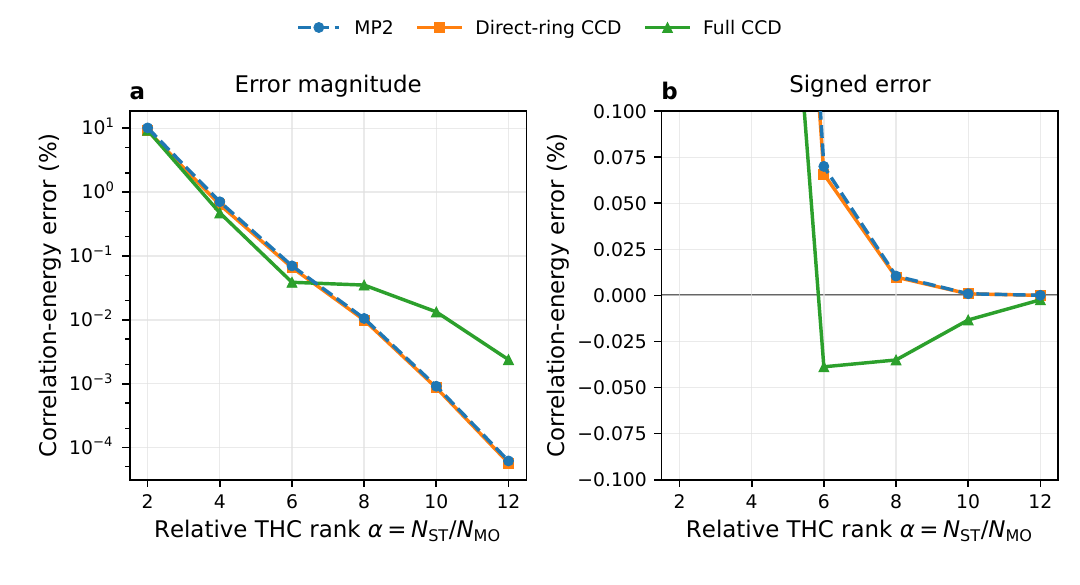}
  \caption{Absolute (a) and signed (b) relative errors in the correlation energy for MP2, drCCD+SOSEX, and CCD calculations of butane in percent as a function of the relative rank $\alpha=N_{\mathrm{ST}}/N_{\mathrm{MO}}$.}
  \label{fig:channels}
\end{figure}

As shown in \cref{fig:channels}, MP2 and drCCD+SOSEX exhibit similar convergence behavior with increasing rank. This suggests that the inclusion of direct ring contributions does not substantially alter the compressibility of the doubles amplitudes, which is consistent with the particle-hole excitation structure of the direct ring terms. In contrast, the slower convergence and the change in the sign of the error observed for full CCD suggest that contributions beyond the direct ring approximation, particularly exchange and ladder terms, may be responsible for the different convergence behavior.

\subsection{Accuracy of Relative Energies}\label{sec:rel_energies}

Most quantum chemical calculations benefit from error cancellation effects, that are only visible when calculating energy differences.
These are relevant, since they describe observable energy changes due to changes in the configuration.
To test this effect for the CCSD energies, obtained in the imaginary-time resolved pair-density basis, we perform a torsional scan of the butane molecule.
The dihedral rotation angle along the bond of the middle two carbon atoms is modified from $0^\circ$ to $360^\circ$ in $15^\circ$ steps.
The configuration at $0^\circ$ is lowest in energy and taken as the zero-point:
\begin{equation}
  \Delta E(\theta)=E(\theta)-E(0^\circ).
\end{equation}
The error of this energy difference, caused by using the imaginary-time resolved interpolation grid is therefore given by
\begin{equation}
 \begin{aligned}
 \delta\Delta E(\theta)={}&
 \left[E^\mathrm{THC}_{\mathrm{CCSD}}(\theta)-E^\mathrm{THC}_{\mathrm{CCSD}}(0^\circ)\right] \\
 &-\left[E_{\mathrm{CCSD}}(\theta)-E_{\mathrm{CCSD}}(0^\circ)\right].
 \end{aligned}
 \label{eq:double-difference-but}
\end{equation}

\begin{figure}[htbp]
  \centering
  \includegraphics[width=0.88\textwidth]{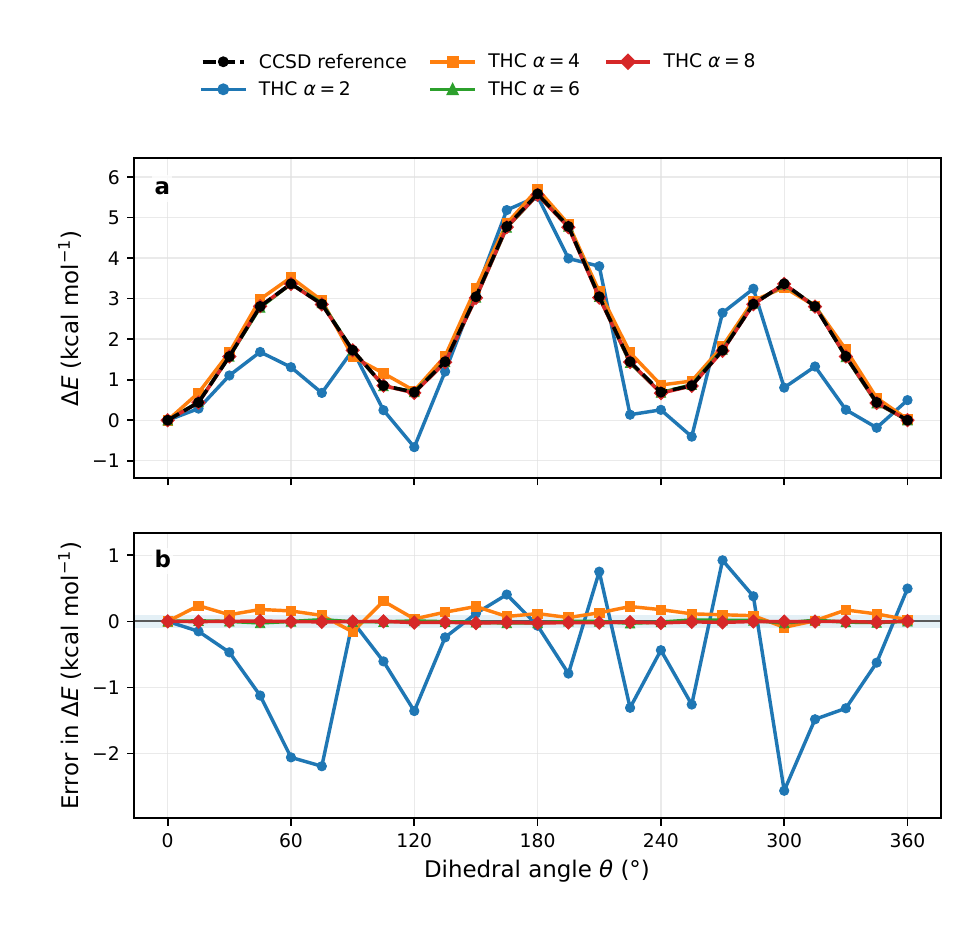}
  \caption{Relative CCSD and THC-CCSD energies of butane as a function of the dihedral angle
  (a) and signed error in the relative energy (b). The
  0$^\circ$ geometry defines zero. The blue region marks errors below $0.1$ kcal mol$^{-1}$.}
  \label{fig:butane-profile}
\end{figure}

Figure \ref{fig:butane-profile} shows the CCSD energy differences calculated in
the imaginary-time resolved interpolation basis at various THC ranks,
as well as the error
that is made with respect to the reference calculation.
The reference energy differences can be reproduced with an accuracy
below $0.1$ kcal/mol by using a relative THC rank of $\alpha = 6$. The results indicate that our CCSD amplitude approximation exhibits error cancellation
for this application, when comparing it to the accuracy of the total energy, presented in Figure \ref{fig:size-convergence} in the previous section. At a THC rank of e.g.
$\alpha=4$, the total CCSD error is still above $2$ kcal mol$^{-1}$, whereas the error in the energy difference shown in \cref{fig:butane-profile}(b) is about one order of magnitude smaller.

As a second test we calculated the dissociation curves of 8 bipartite systems taken from the S66x8 data set.
For the energies differences we chose the energy of the configurations where the molecules are the furthest apart as the zero-point:
\begin{equation}
  \Delta E(R)=E(R)-E(2R_e).
\end{equation}
The accuracy-defining quantity is the error of this difference:
\begin{equation}
 \begin{aligned}
 \delta\Delta E(R)={}&
 \left[E^\mathrm{THC}_{\mathrm{CCSD}}(R)-E^\mathrm{THC}_{\mathrm{CCSD}}(2R_e)\right] \\
 &-\left[E_{\mathrm{CCSD}}(R)-E_{\mathrm{CCSD}}(2R_e)\right].
 \end{aligned}
 \label{eq:double-difference}
\end{equation}
\begin{figure}[htbp]
  \centering
  \includegraphics[width=0.92\textwidth]{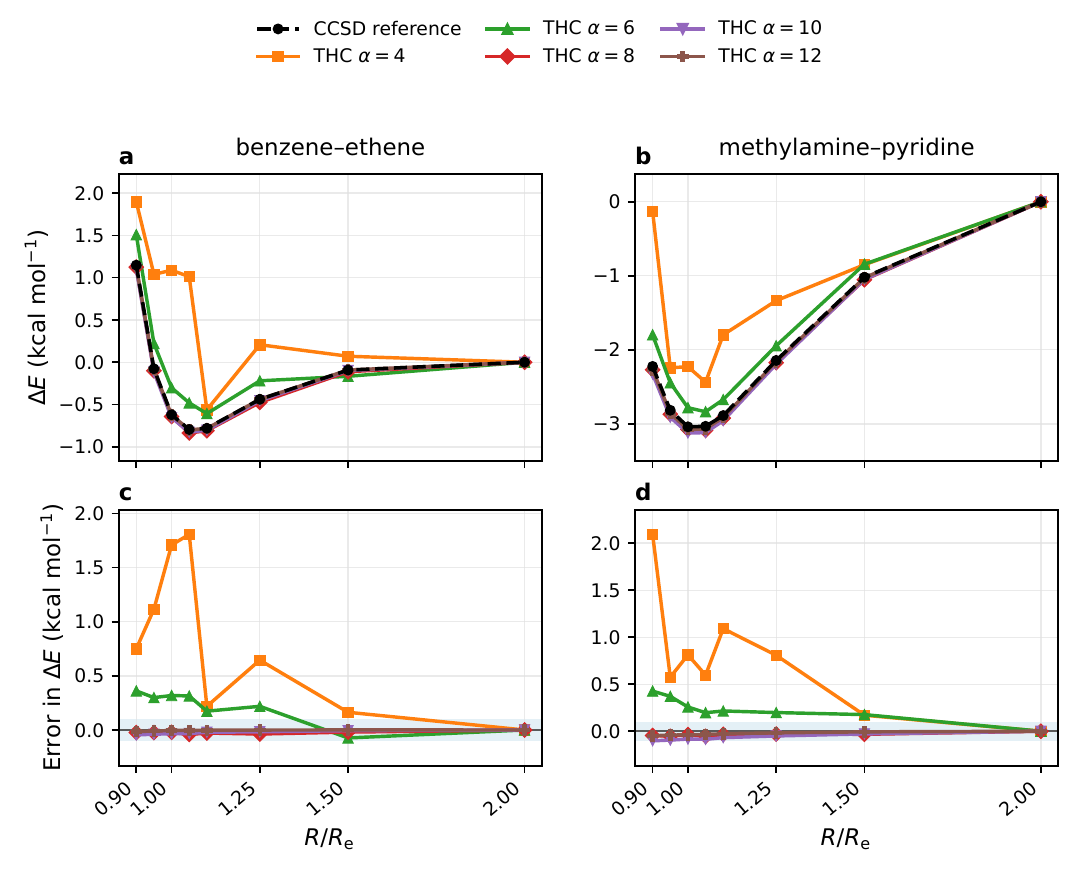}
  \caption{Side-by-side comparison of the dissociation curve and the error made by the THC approximation of the doubles amplitudes of benzene-ethene (a),(c) and methylamine-pyridine (b),(d).
  Energies are referenced to the geometry where the molecules are furthest apart, $2R_e$. The blue region marks errors below $0.1$ kcal mol$^{-1}$.}
  \label{fig:benz-eth_menh2-pyr}
\end{figure}

\cref{fig:benz-eth_menh2-pyr} shows the dissociation curves of the two largest systems in the test set for several sizes of the imaginary-time-resolved interpolation grid. At a relative THC rank of $\alpha=10$, the energy differences between the equilibrium configuration and the configuration with the largest intermolecular separation are converged to within $0.1$ kcal mol$^{-1}$. The dissociation curves for all eight systems in S66$\times$8/8 are provided in the supplementary material. For some of the smaller systems, such as water--water, calculations at such high THC ranks were not performed because the resulting representation would be larger than the full CCSD doubles-amplitude tensor and thus provide no compression benefit.

Across all systems, slightly fewer grid points are required to achieve a given accuracy at larger intermolecular separations. This trend is consistent with the grid construction: the point-selection metric depends on real-space orbital overlaps, and the overlap between orbitals on different subsystems decreases as their separation increases.

\section{Conclusions}\label{sec:conclusions}

We introduce an approximation to the propagated orbital-pair densities based on an interpolation grid resolved in both imaginary time and real space. We demonstrate that this grid provides a compact representation of the doubles amplitudes used by the second-order M{\o}ller-Plesset theory and can also be used in rank-reduced coupled-cluster singles and doubles (CCSD) calculations.

The energy errors exhibit two convergence regimes: a rapid initial decrease up to a relative THC basis size of $\alpha\approx 4$–$6$, followed by slower convergence. This behavior is consistent with the efficient representation of the MP2 contribution to the CCSD doubles amplitudes and the slightly less efficient representation of contributions beyond MP2. For all systems considered, $\alpha=8$ yields energy errors below $0.1\%$ of the reference energy. The butane torsional energy profile is converged to within $0.1$ kcal mol$^{-1}$ at $\alpha=6$. For the two largest intermolecular systems, the energy differences between the equilibrium configuration and the configuration with the largest separation are converged to within $0.1$ kcal mol$^{-1}$ at $\alpha=10$. 

The grid also exhibits sensitivity to spatial locality. Across all intermolecular systems, slightly fewer grid points are required to achieve a given accuracy at larger separations. This trend is consistent with the overlap-based selection metric adapting to the reduced overlap between orbitals on different subsystems.

The present CCSD implementation serves to assess the proposed compression and does not yet reduce computational scaling. In principle, the compressed formulation permits $\mathcal{O}(N^4)$ scaling.  An efficient implementation must also balance the cost of lower-scaling terms with large prefactors against that of higher-scaling terms with small prefactors. Further algorithmic development is therefore required to translate the compact representation into computational savings. The separability of the orbital and $k$-point indices in the THC representation may facilitate an extension to periodic systems. In addition, the imaginary-time-resolved construction of the basis motivates investigating its application to finite-temperature CCSD and related methods. Future work will focus on an optimized implementation, these extensions, and broader numerical validation.

\begin{acknowledgement}
This work was supported by the Flatiron Institute and the Simons Foundation. We also acknowledge the Scientific Computing Core (SCC) at Flatiron Institute\\ (https://www.simonsfoundation.org/flatiron/scientific-computing-core/) for providing computational resources and technical support that have contributed to the results reported within this paper. 
\end{acknowledgement}

\bibliography{kmprefs,extra_l_isdf}

\clearpage

\section*{Supporting Information}
\setcounter{figure}{0}
\setcounter{table}{0}
\setcounter{equation}{0}
\renewcommand{\thefigure}{S\arabic{figure}}
\renewcommand{\thetable}{S\arabic{table}}
\renewcommand{\theequation}{S\arabic{equation}}

\providecommand{\theHfigure}{}
\providecommand{\theHtable}{}
\providecommand{\theHequation}{}
\renewcommand{\theHfigure}{SI.\arabic{figure}}
\renewcommand{\theHtable}{SI.\arabic{table}}
\renewcommand{\theHequation}{SI.\arabic{equation}}

Calculations used the settings described in \cref{sec:details} of the main text. 

\subsection*{Accuracy of Total Energies}

\begin{figure}[htbp]
\centering
\includegraphics[width=0.84\textwidth]{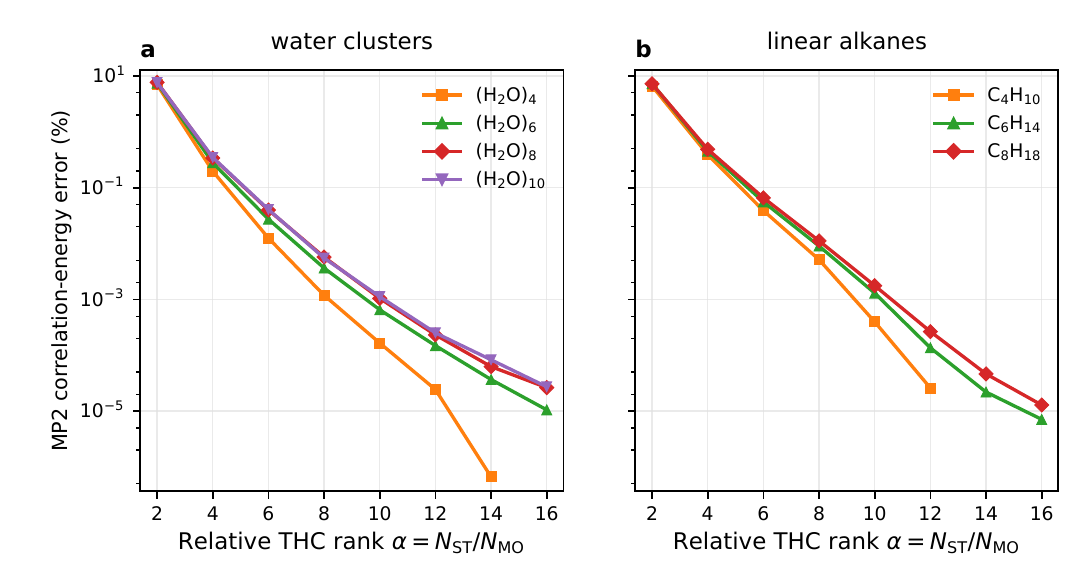}
\caption{Absolute error in the total MP2 correlation energy in percent as a function
  of $\alpha=N_\mathrm{ST}/N_{\mathrm{MO}}$ for water clusters (a) and linear alkanes (b).}
\end{figure}

\begin{figure}[htbp]
\centering
\includegraphics[width=0.84\textwidth]{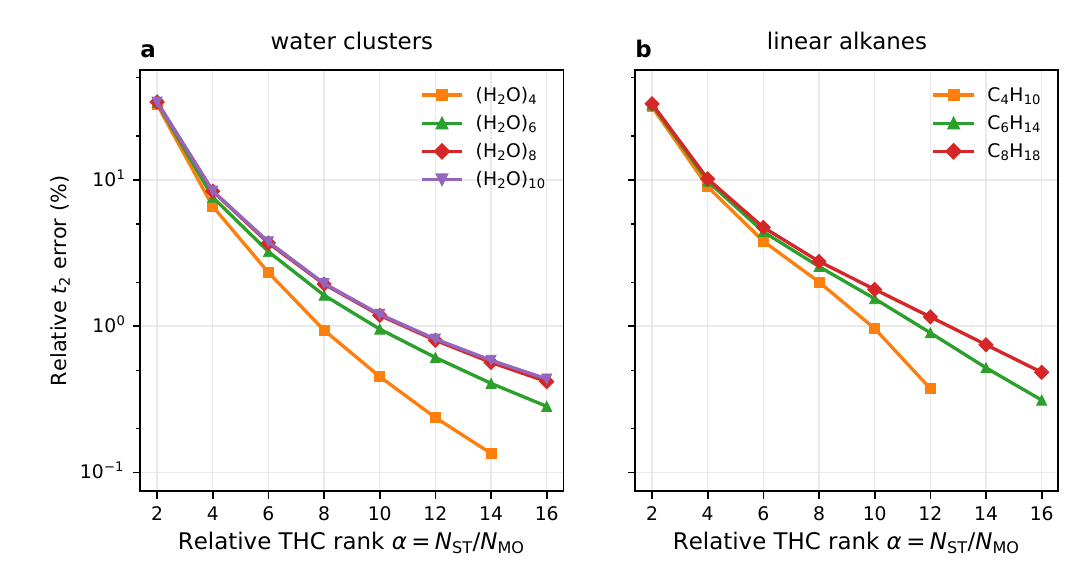}
\caption{$l^2$ error of the THC-rank-reduced $t_2$ amplitude relative to the full $t_2$ amplitudes as a function
  of $\alpha=N_\mathrm{ST}/N_{\mathrm{MO}}$ for water clusters (a) and linear alkanes (b).}
\end{figure}
\newpage
\subsection*{Accuracy of Relative Energies -- System-resolved S66x8 curves}

\begin{figure}[htbp]
\centering
\includegraphics[width=0.48\linewidth]{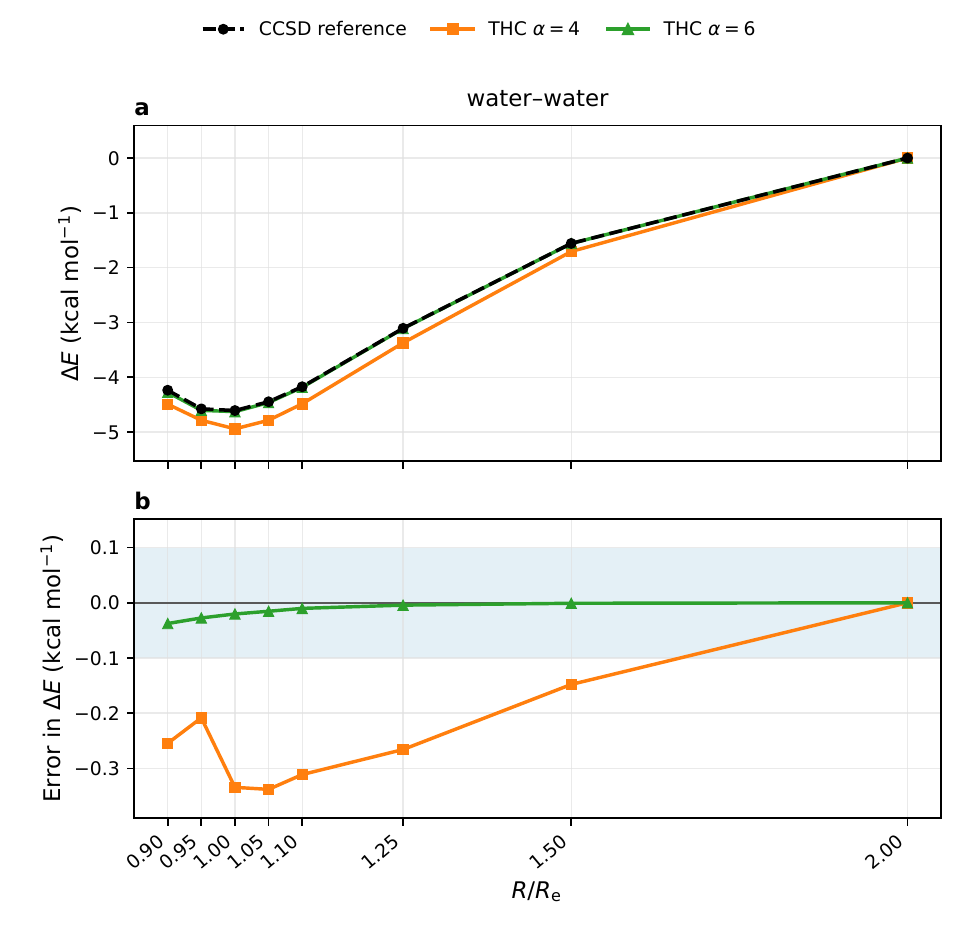}
\hfill
\includegraphics[width=0.48\linewidth]{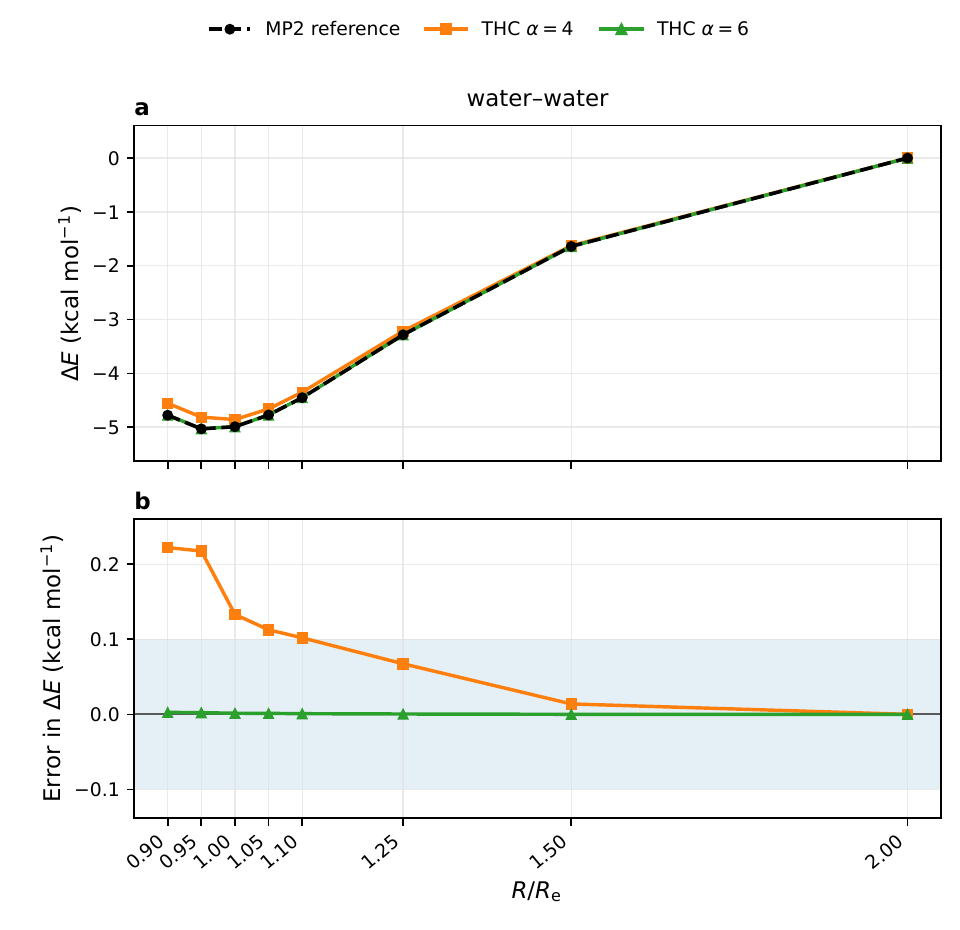}
\caption{Side-by-side comparison of the dissociation curve and the error made by the THC approximation of the doubles amplitudes of water--water calculated at the CCSD (left) and MP2 (right) level.
  Energies are referenced to the geometry where the molecules are furthest apart, $2R_e$. The blue region marks errors below $0.1$ kcal mol$^{-1}$.}
\end{figure}

\begin{figure}[htbp]
\centering
\includegraphics[width=0.48\linewidth]{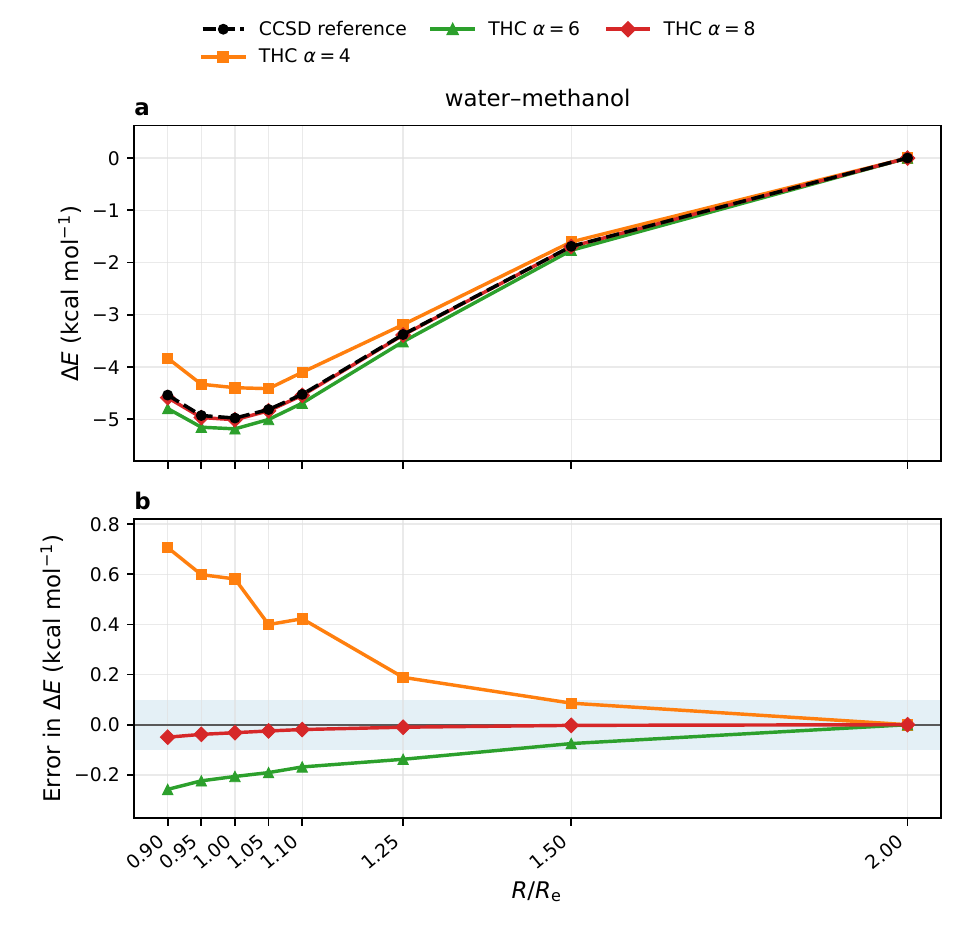}
\hfill
\includegraphics[width=0.48\linewidth]{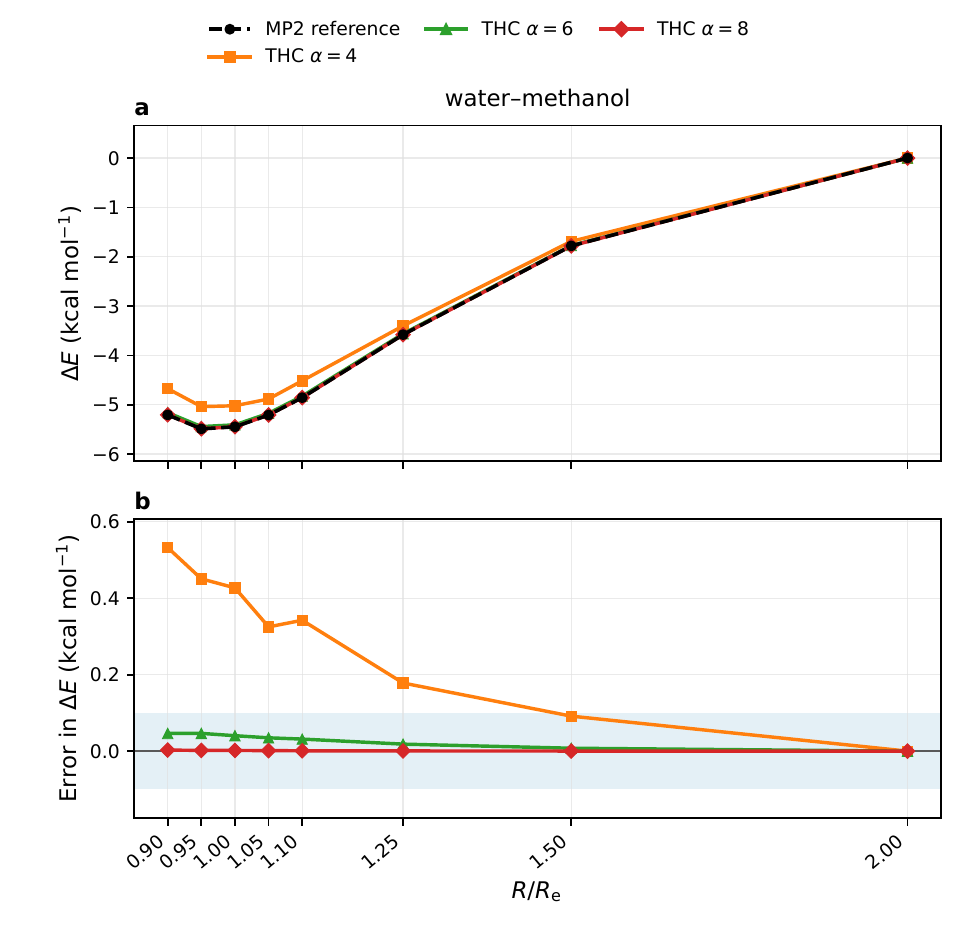}
\caption{Side-by-side comparison of the dissociation curve and the error made by the THC approximation of the doubles amplitudes of water--methanol calculated at the CCSD (left) and MP2 (right) level.
  Energies are referenced to the geometry where the molecules are furthest apart, $2R_e$. The blue region marks errors below $0.1$ kcal mol$^{-1}$}
\end{figure}

\begin{figure}[htbp]
\centering
\includegraphics[width=0.48\linewidth]{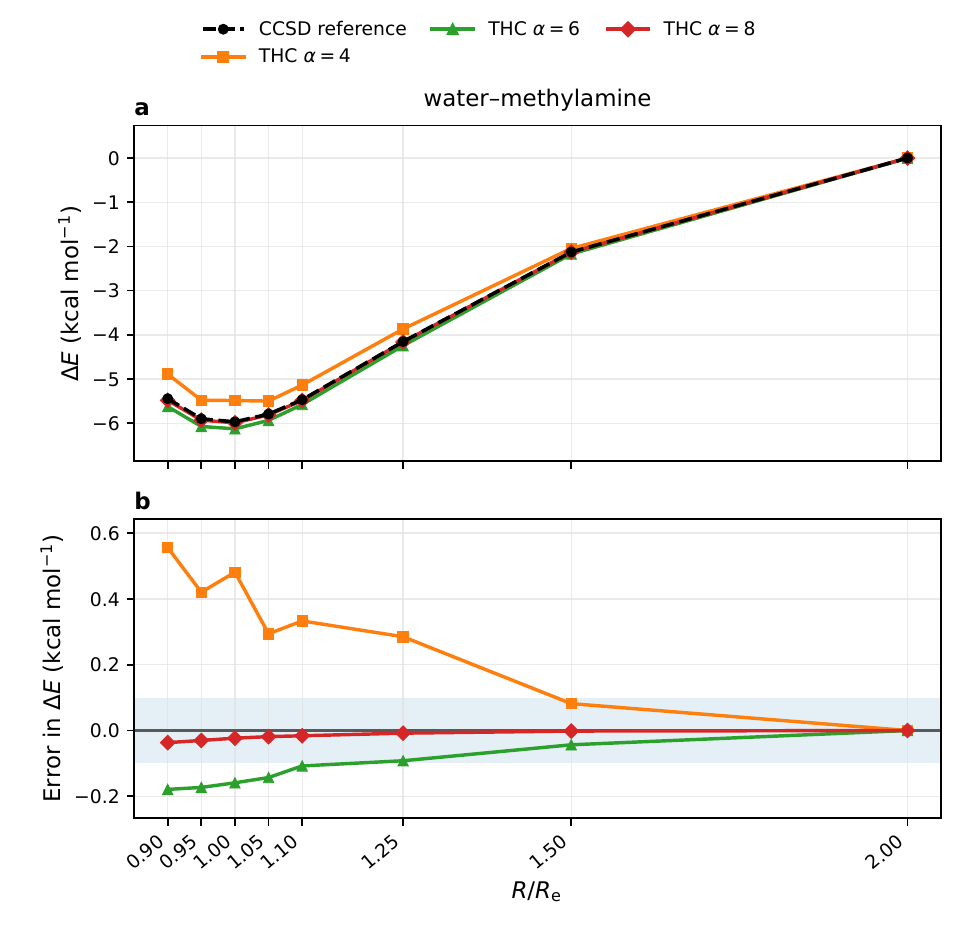}
\hfill
\includegraphics[width=0.48\linewidth]{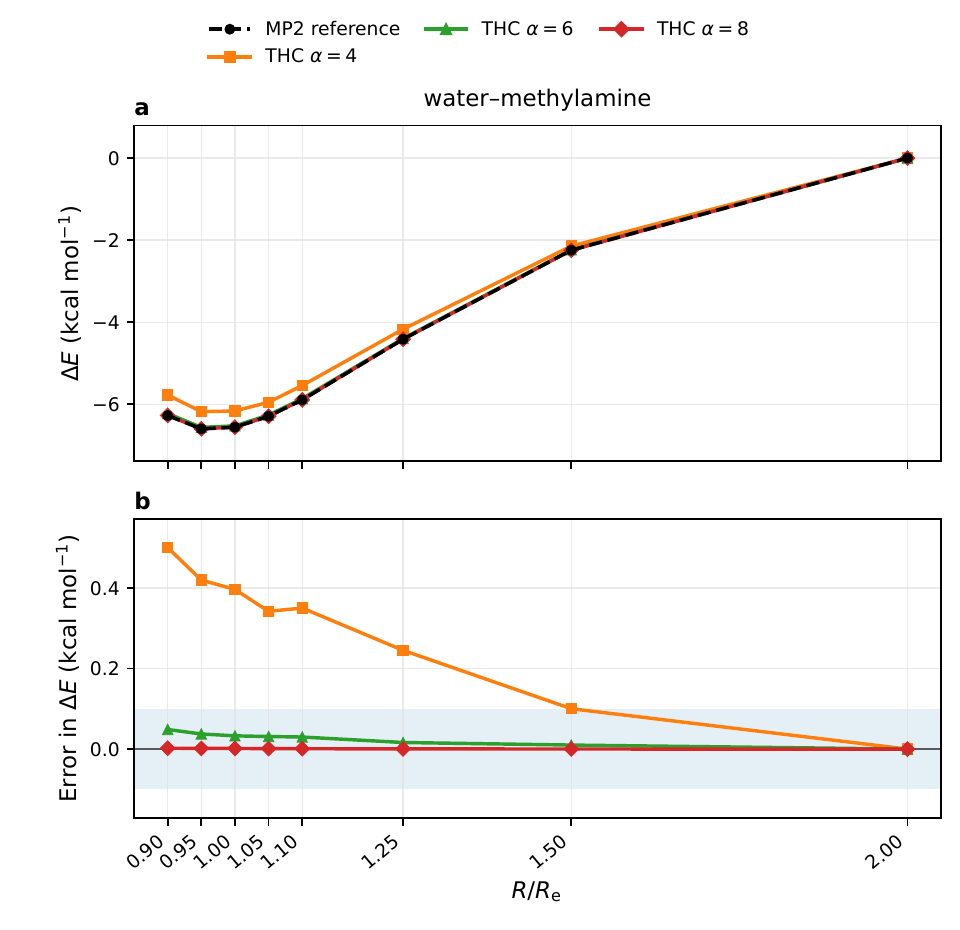}
\caption{Side-by-side comparison of the dissociation curve and the error made by the THC approximation of the doubles amplitudes of water--methylamine calculated at the CCSD (left) and MP2 (right) level.
  Energies are referenced to the geometry where the molecules are furthest apart, $2R_e$. The blue region marks errors below $0.1$ kcal mol$^{-1}$.}
\end{figure}

\begin{figure}[htbp]
\centering
\includegraphics[width=0.48\linewidth]{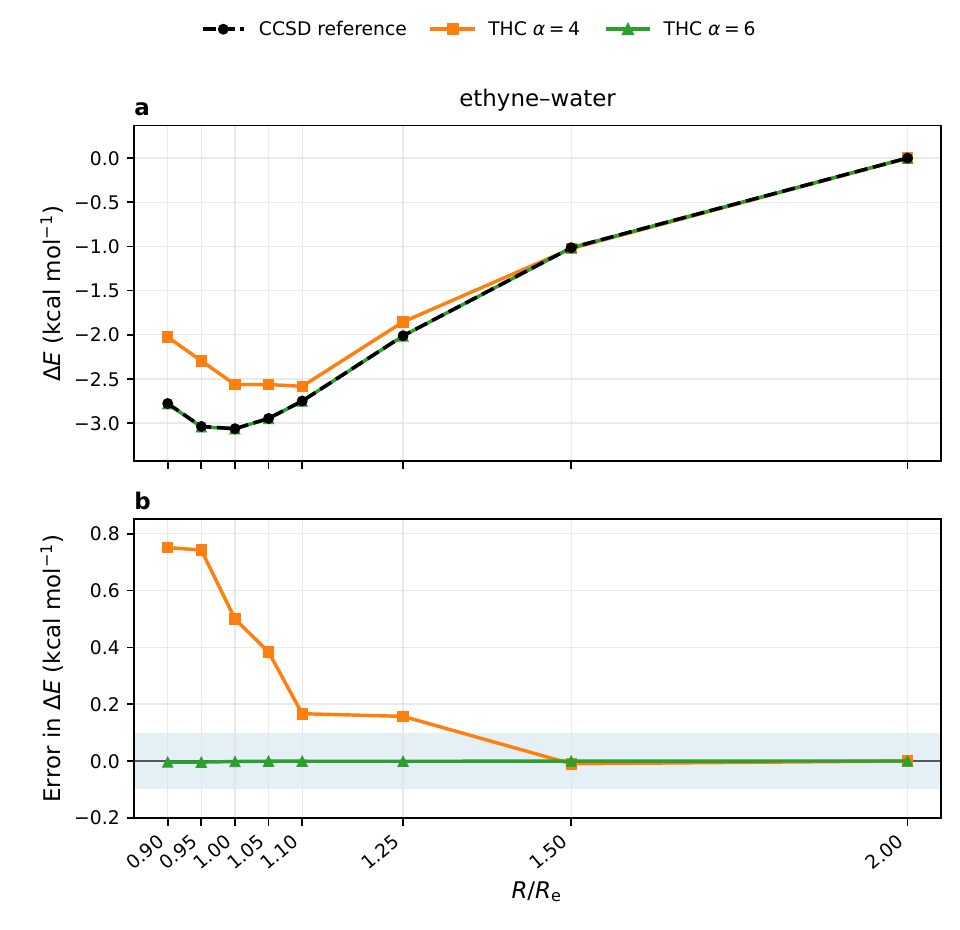}
\hfill
\includegraphics[width=0.48\linewidth]{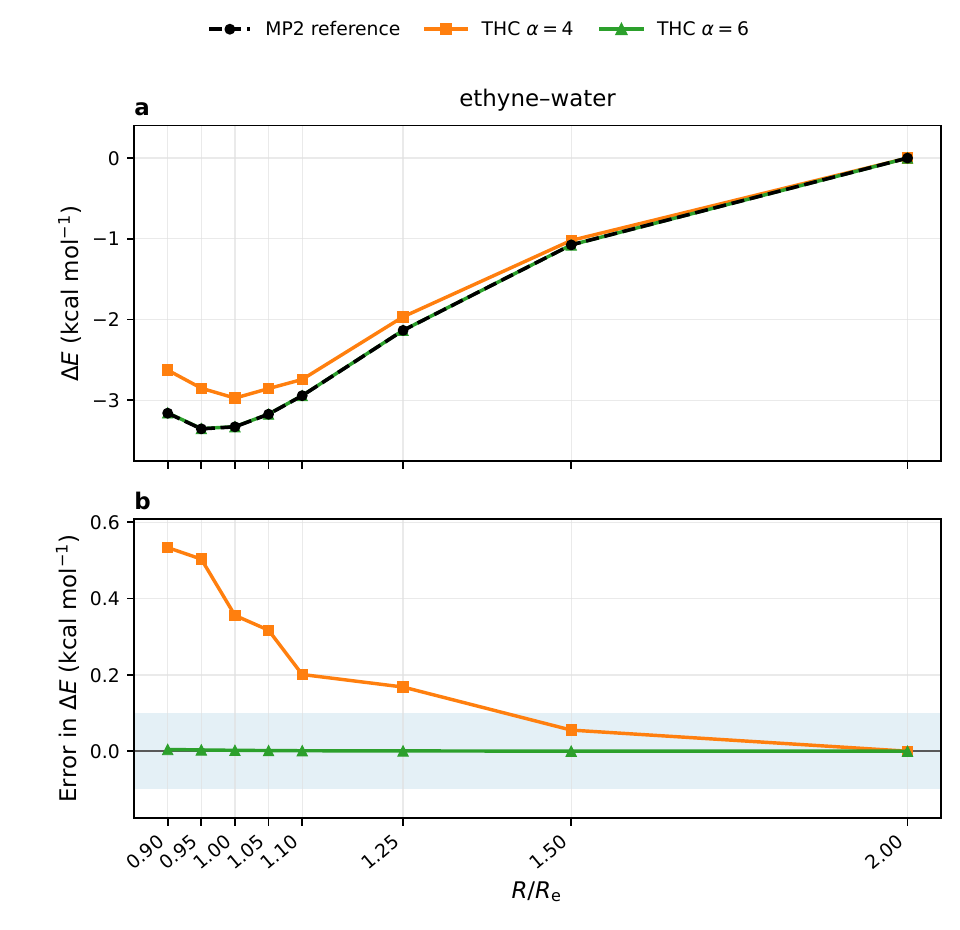}
\caption{Side-by-side comparison of the dissociation curve and the error made by the THC approximation of the doubles amplitudes of ethyne--water calculated at the CCSD (left) and MP2 (right) level.
  Energies are referenced to the geometry where the molecules are furthest apart, $2R_e$. The blue region marks errors below $0.1$ kcal mol$^{-1}$.}
\end{figure}

\begin{figure}[htbp]
\centering
\includegraphics[width=0.48\linewidth]{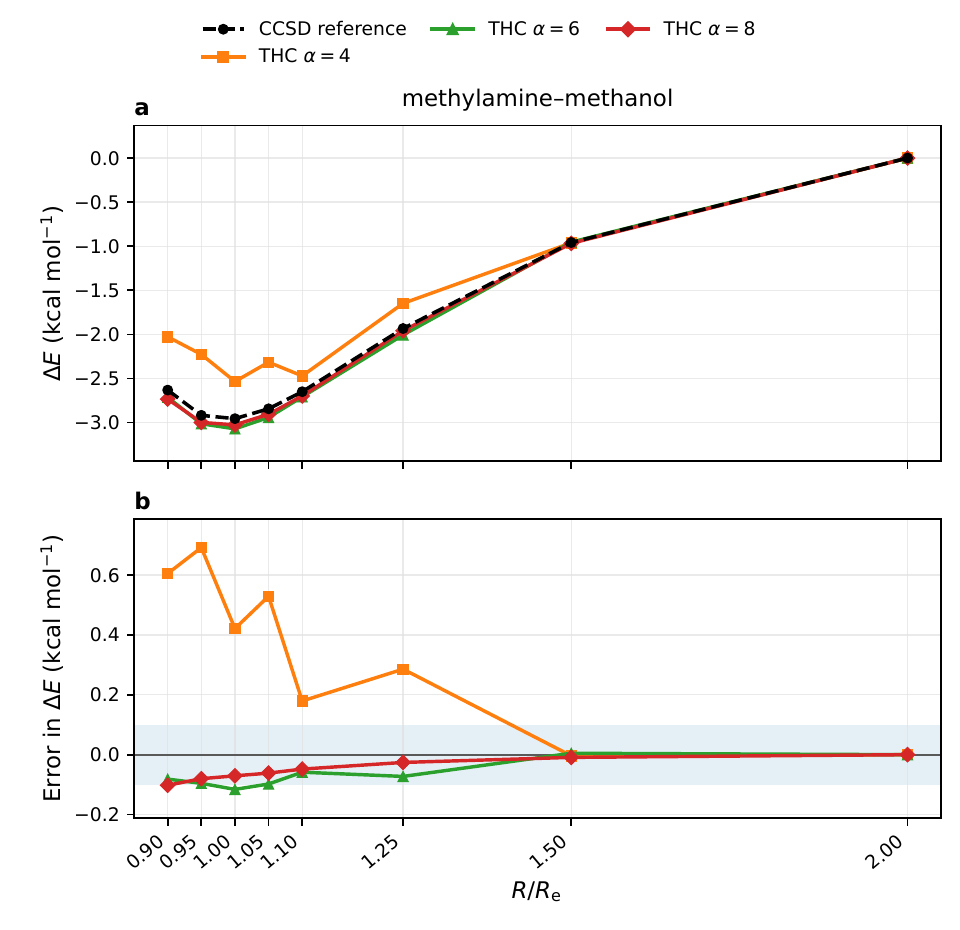}
\hfill
\includegraphics[width=0.48\linewidth]{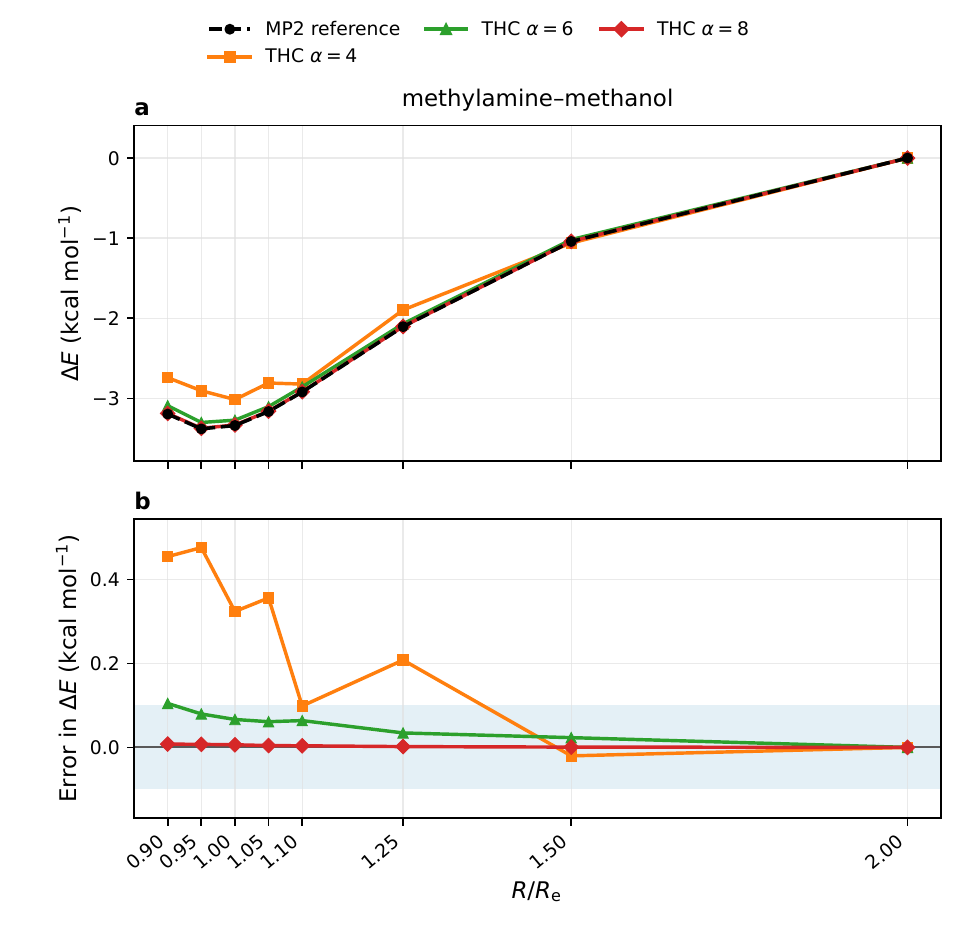}
\caption{Side-by-side comparison of the dissociation curve and the error made by the THC approximation of the doubles amplitudes of methylamine--methanol calculated at the CCSD (left) and MP2 (right) level.
  Energies are referenced to the geometry where the molecules are furthest apart, $2R_e$. The blue region marks errors below $0.1$ kcal mol$^{-1}$.}
\end{figure}

\begin{figure}[htbp]
\centering
\includegraphics[width=0.48\linewidth]{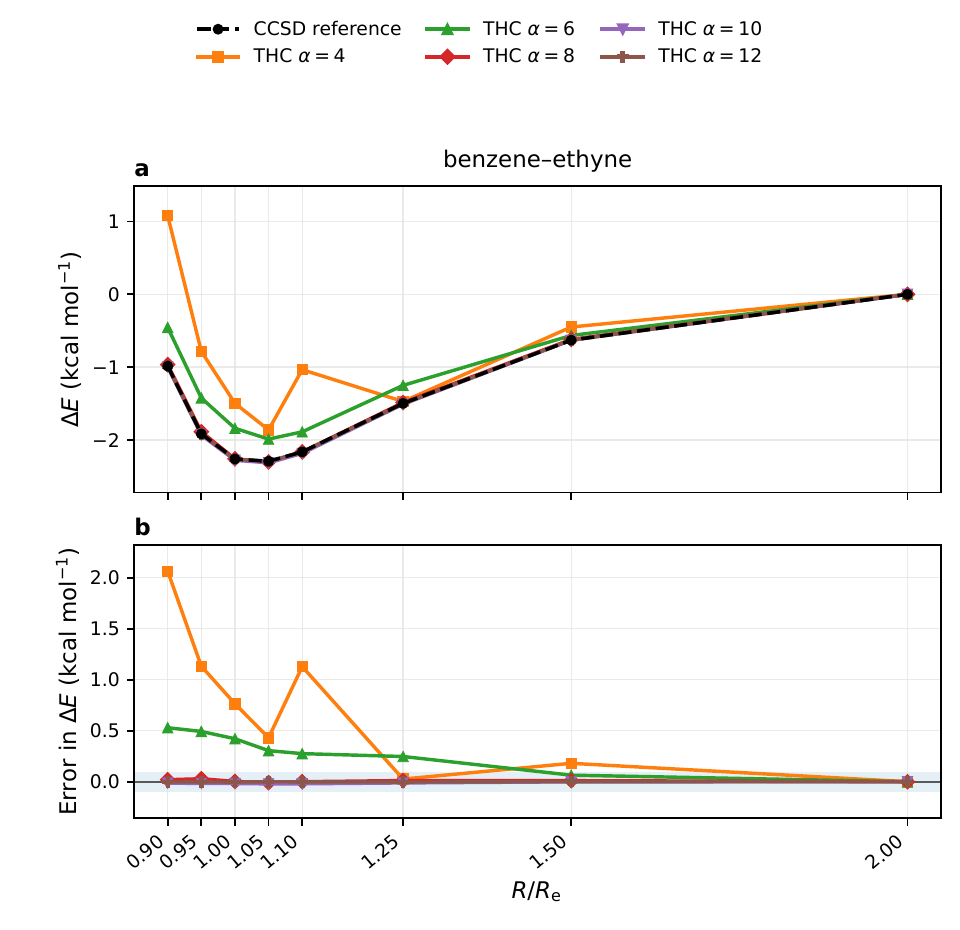}
\hfill
\includegraphics[width=0.48\linewidth]{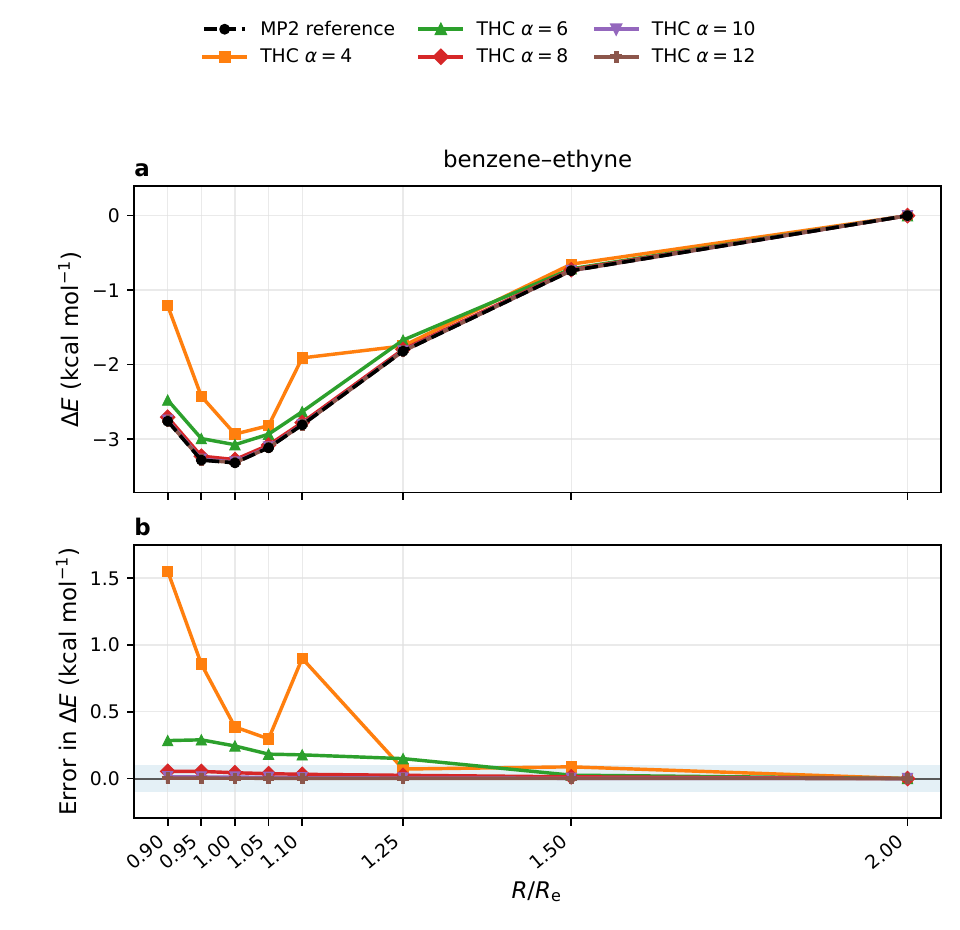}
\caption{Side-by-side comparison of the dissociation curve and the error made by the THC approximation of the doubles amplitudes of benzene--ethyne calculated at the CCSD (left) and MP2 (right) level.
  Energies are referenced to the geometry where the molecules are furthest apart, $2R_e$. The blue region marks errors below $0.1$ kcal mol$^{-1}$.}
\end{figure}

\begin{figure}[htbp]
\centering
\includegraphics[width=0.48\linewidth]{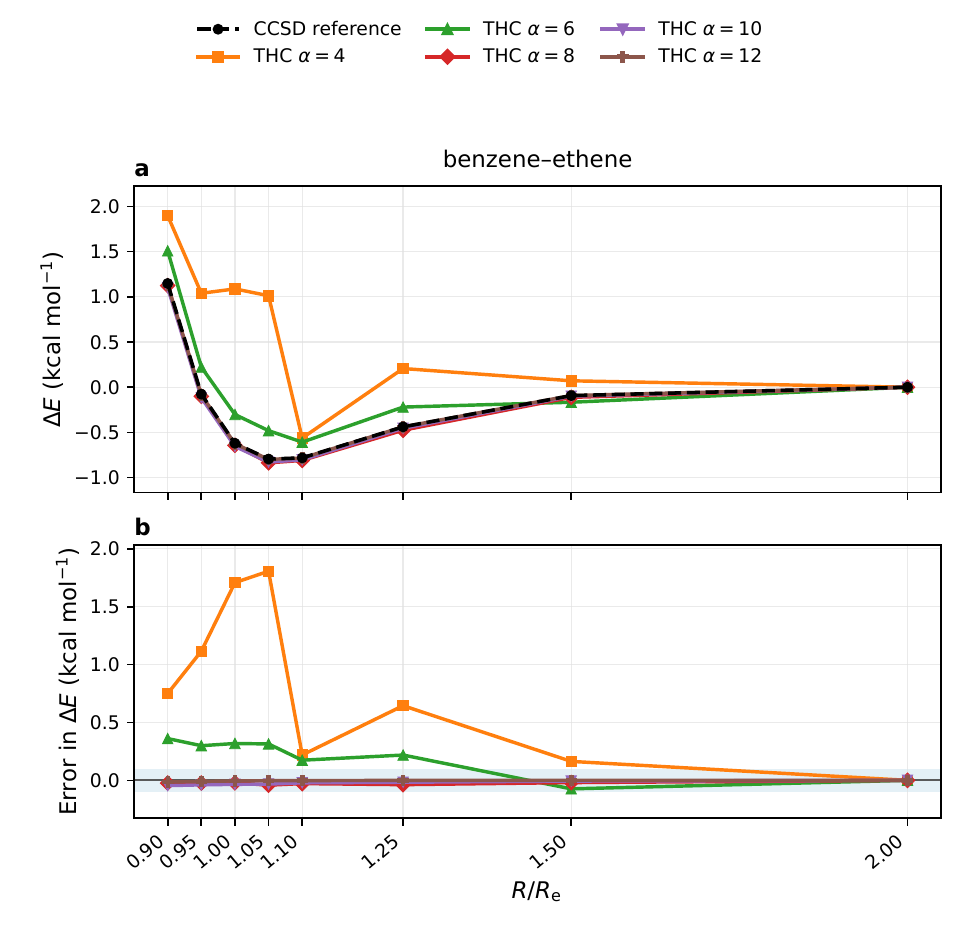}
\hfill
\includegraphics[width=0.48\linewidth]{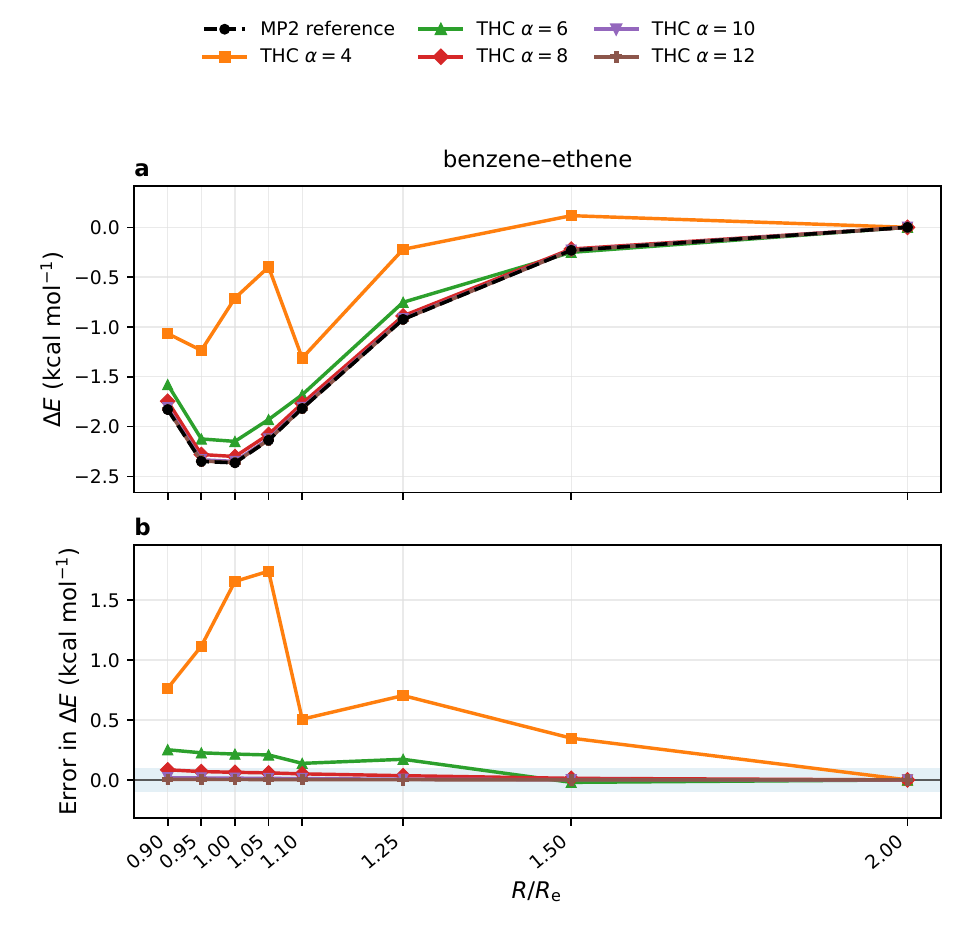}
\caption{Side-by-side comparison of the dissociation curve and the error made by the THC approximation of the doubles amplitudes of benzene--ethene calculated at the CCSD (left) and MP2 (right) level.
  Energies are referenced to the geometry where the molecules are furthest apart, $2R_e$. The blue region marks errors below $0.1$ kcal mol$^{-1}$.}
\end{figure}

\begin{figure}[htbp]
\centering
\includegraphics[width=0.48\linewidth]{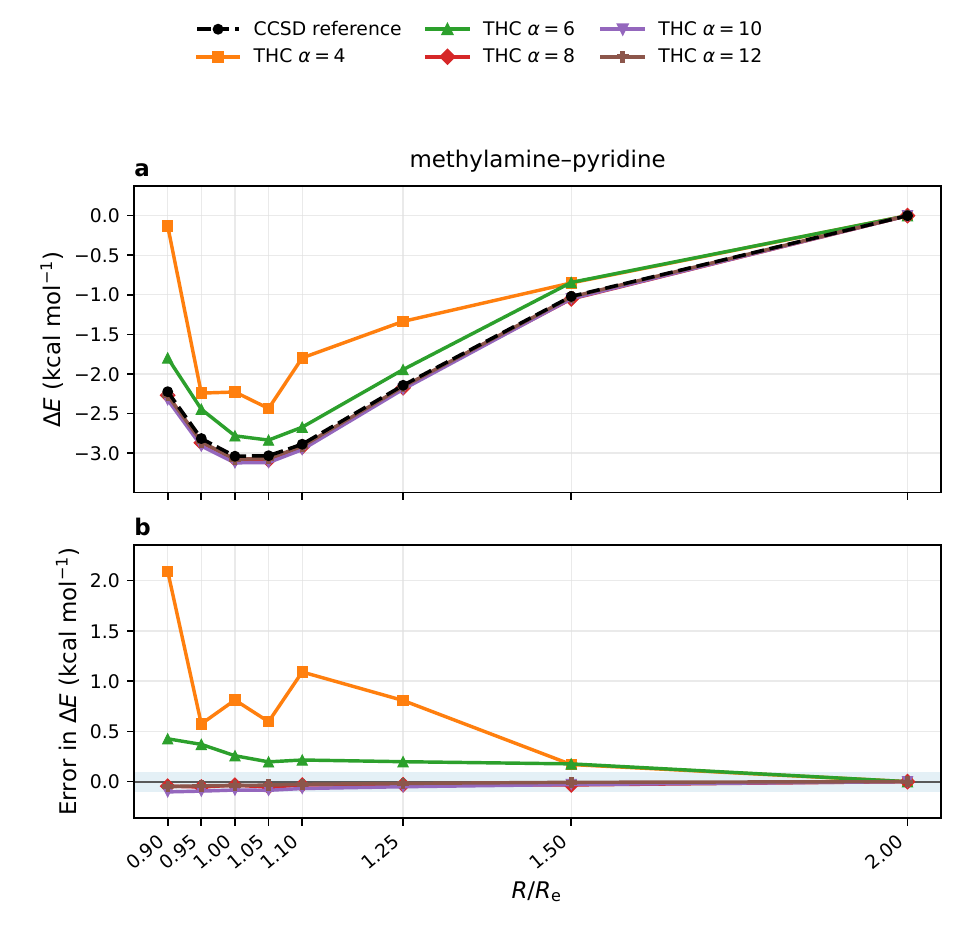}
\hfill
\includegraphics[width=0.48\linewidth]{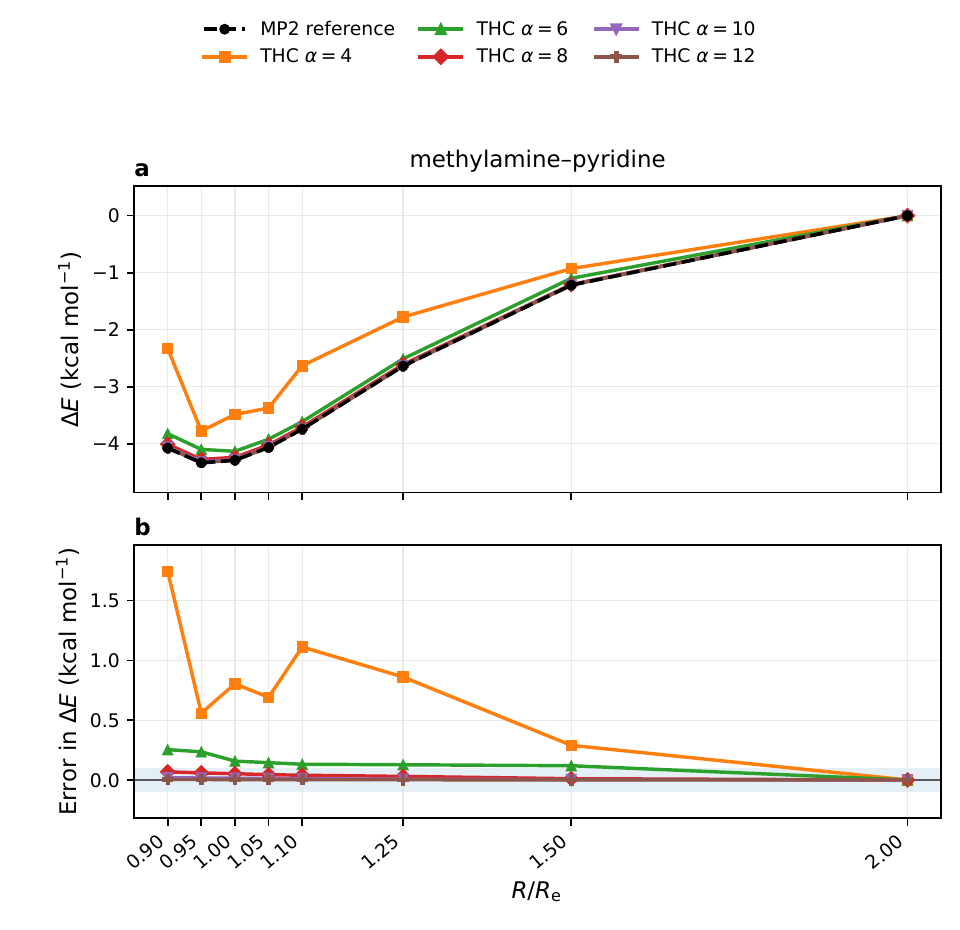}
\caption{Side-by-side comparison of the dissociation curve and the error made by the THC approximation of the doubles amplitudes of methylamine--pyridine calculated at the CCSD (left) and MP2 (right) level.
  Energies are referenced to the geometry where the molecules are furthest apart, $2R_e$. The blue region marks errors below $0.1$ kcal mol$^{-1}$.}
\end{figure}

\end{document}